\documentclass[a4paper,11pt]{article}
\usepackage{jheppub} 

\preprint{INT-PUB-25-019, YITP-SB-2025-11}
\title{Graph theory inspired anomaly detection at the LHC}

\usepackage[utf8]{inputenc}
\usepackage[T1]{fontenc}
\usepackage{lmodern}
\usepackage{microtype}
\usepackage{mathtools}
\usepackage{amsmath}

\usepackage{graphicx,hyperref,braket,orcidlink, algorithm2e, amssymb, amsmath, dsfont,subfigure,multirow}

\usepackage{placeins}

\usepackage{amsfonts}
\usepackage{bm}
\usepackage{graphicx}
\usepackage{mwe}
\usepackage{multirow}
\hypersetup{
  colorlinks=true,
  citecolor=blue,
  linkcolor=blue,
  urlcolor=blue
}
\usepackage{sidecap}

\usepackage{color}
\definecolor{darkblue}{rgb}{0,0,0.5}

\author[1,2,3]{Jack Y. Araz,}
\emailAdd{j.araz@ucl.ac.uk}

\author[3,4]{Dimitrios Athanasakos,}
\emailAdd{dimitrios.athanasakos@stonybrook.edu}

\author[5]{Mateusz Ploskon,}
\emailAdd{mploskon@lbl.gov}

\author[3]{Felix Ringer}
\emailAdd{felix.ringer@stonybrook.edu}

\affiliation[1]{Department of Physics and Astronomy, University College London, London, WC1E 6B, UK}
\affiliation[2]{School of Science \& Technology, City St. George’s, University of London, London, EC1V0HB, UK}
\affiliation[3]{Department of Physics and Astronomy, Stony Brook University, Stony Brook, NY 11794, USA}
\affiliation[4]{C.N. Yang Institute for Theoretical Physics, Stony Brook University, Stony Brook, NY 11794, USA}
\affiliation[5]{Nuclear Science Division, Lawrence Berkeley National Laboratory, Berkeley, CA 94720,
USA}

\abstract{Designing model-independent anomaly detection algorithms for analyzing LHC data remains a central challenge in the search for new physics, due to the high dimensionality of collider events. In this work, we develop a graph autoencoder as an unsupervised, model-agnostic tool for anomaly detection, using the LHC Olympics dataset as a benchmark. By representing jet constituents as a graph, we introduce a method to systematically control the information available to the model through sparse graph constructions that serve as physically motivated inductive biases. Specifically, (1) we construct graph autoencoders based on locally rigid Laman graphs and globally rigid unique graphs, and (2) we explore the clustering of jet constituents into subjets to interpolate between high- and low-level input representations. We obtain the best performance, measured in terms of the Significance Improvement Characteristic curve for an intermediate level of subjet clustering and certain sparse unique graph constructions. We further investigate the role of graph connectivity in jet classification tasks. Our results demonstrate the potential of leveraging graph-theoretic insights to refine and increase the interpretability of machine learning tools for collider experiments.}

\begin{document}

\maketitle

\section{Introduction}
\label{section:intro}

Decades of experimental validation,  have confirmed Standard Model's (SM) predictions with remarkable precision. However, the SM is widely understood to be an incomplete theory. It does not incorporate gravity, or account for the strong evidence of dark matter and dark energy. These and other issues serve as compelling motivations for the existence of physics beyond the Standard Model (BSM). Since the Higgs discovery~\cite{CMS:2012qbp, ATLAS:2012yve}, the search for BSM signals is one of the central goals of the experimental program at the LHC.  Historically, these searches have largely followed a ``top-down'' approach, targeting specific theoretical models proposed to address the SM's shortcomings ~\cite{CMS:2019zmd, CMS:2019zmn, ATLAS:2016gty, CMS:2016ohy, ATLAS:2017nga, ATLAS:2016bek}. While this targeted search strategy has explored vast regions of potential new physics parameter space, it inherently carries the risk of being blind to signatures that do not align closely with the specific models being tested.

Given the absence of a definitive BSM discovery thus far, despite extensive searches, there is a growing interest in complementary search strategies that are essential to take advantage of the large data sets from the LHC. This has led to significant efforts in model-agnostic approaches using machine learning, as can be seen for example from the 2020 LHC Olympics anomaly detection challenge~\cite{lhco, kasieczka_2019_4536624}. These approaches aim to identify potential new physics signals by detecting anomalous events from the Standard Model background. While less sensitive to the signatures of any particular model, they enable simultaneous searches across a wide range of BSM scenarios in a theory-agnostic way. Moreover, they allow training directly on real, unlabeled data instead of relying only on simulations~\cite{Gambhir:2025afb, Aarrestad:2021oeb, ATLAS:2025obc, Knapp:2020dde, ATLAS:2023ixc,Bhardwaj:2024djv}. A comprehensive overview of recent developments in anomaly detection methods at particle colliders can be found in Ref.~\cite{Belis:2023mqs}.

Most machine learning–based anomaly detection algorithms fall into two main categories: approaches based on unsupervised or weakly supervised training. Within the unsupervised paradigm, various studies have been performed using autoencoders~\cite{Kingma:2013hel, Cerri:2018anq, Cheng:2020dal, Dillon:2021nxw, Heimel:2018mkt, Farina:2018fyg, Bortolato:2021zic, Vaslin:2023lig, Jawahar:2021vyu, Finke:2021sdf, Laguarta:2023evo, Blance:2019ibf, Dillon:2022mkq, Schuhmacher:2023pro}. A standard autoencoder consists of two neural networks: the encoder and the decoder, that are trained to reconstruct the input. Due to an intermediate bottleneck, the autoencoder is forced to learn the dominant features of the data distribution. In collider physics applications, an autoencoder can be trained on background-rich data and evaluated on a mixed data set, where events with high reconstruction loss are flagged as anomalous. 

Recently, much work has focused on applying GNNs for autoencoder-based anomaly detection, moving beyond simple dense representations. Ref. \cite{tsan2021particlegraphautoencodersdifferentiable} used graph convolutional networks with an EMD-based \cite{Komiske_2019} reconstruction loss to model jets as particle clouds. To ensure theoretical consistency, \cite{Konar:2021zdg} developed an IRC-safe architecture that uses energy-weighted aggregation to suppress soft radiation. More recently, Geometric Deep Learning \cite{bronstein2021geometricdeeplearninggrids} methods have been applied to enforce Lorentz equivariance in the latent space \cite{Hao_2023}, while Transformer-based models have leveraged global attention mechanisms for improved performance \cite{liu2023}. Our work investigates the impact of using sparse graph structures motivated by graph theory.

Weakly supervised anomaly detection often employs the Classification Without Labels (CWoLa) approach~\cite{Metodiev:2017vrx, Dery:2017fap, Finke:2022lsu, Collins:2019jip, Amram:2020ykb}, particularly for searches where a new signal is localized in a certain phase space region. A classifier is trained to distinguish events in this signal region from those in background-dominated sideband regions. To train classifiers on the low-level full event information, generative models have been developed to interpolate the sideband regions into the signal region. See for example Refs.~\cite{Andreassen:2020nkr, cathode, Buhmann:2023acn, Hallin:2022eoq, cheng2025incorporatingphysicalpriorsweaklysupervised, Golling:2023yjq, Golling:2022nkl, dAgnolo:2021aun}.

There are a few differences between unsupervised and weakly supervised anomaly detection techniques that lead to complementary sensitivity~\cite{Collins:2021nxn}. First, the weakly supervised approach relies on the signal fraction when training the classifier. As a result, the performance can improve significantly with an increasing signal fraction. In contrast, in an unsupervised setting such as the autoencoder used in this work, the machine is trained to reconstruct the input features without using labels, and its performance is independent of the signal-to-background ratio, provided that the training sample is dominated by background events.  It is therefore particularly well suited for anomaly detection in the small signal fraction regime~\cite{Collins:2021nxn}. Second, unsupervised anomaly detection can be readily applied to both resonant and non-resonant anomaly detection~\cite{Fraser_2022, vanbeekveld2020combiningoutlieranalysisalgorithms, Caron_2022, collins2022explorationlearntrepresentationsw, Mikuni:2021nwn,Bradshaw_2022, dillon2023normalizedautoencoderlhctriggers, Roche:2023int, Liu:2023djx}, whereas weakly supervised methods have been developed primarily for resonant searches. That said, weakly supervised methods can also be extended to the non-resonant regime. See, for example, Ref.~\cite{Bai:2023yyy}. Third, autoencoders tend to be sensitive only to particular topologies, while weakly supervised methods are generally more robust across a broader class of signals. In this regard, the two approaches are complementary.

Another axis to categorize anomaly detection algorithms is whether they operate on high-level or low-level features such as $N$-subjettiness observables~\cite{Thaler:2010tr,Stewart:2010tn} or full particle four-momenta, respectively. In this work, we address this aspect by clustering jet constituents into subjets using the exclusive $k_T$ algorithm~\cite{Catani:1993hr}. This enables us to interpolate smoothly between using a small number of subjets as high-level inputs and using individual hadrons as low-level inputs to the learning algorithm. This approach removes the need to hand-select a few physically motivated observables. Indeed, we find that the performance of our anomaly detection algorithm peaks for an intermediate number of reclustered subjets, around $n_{\rm subjets}\approx 30$.

\begin{figure*}[t]
\centering
\includegraphics[width=0.32\textwidth]{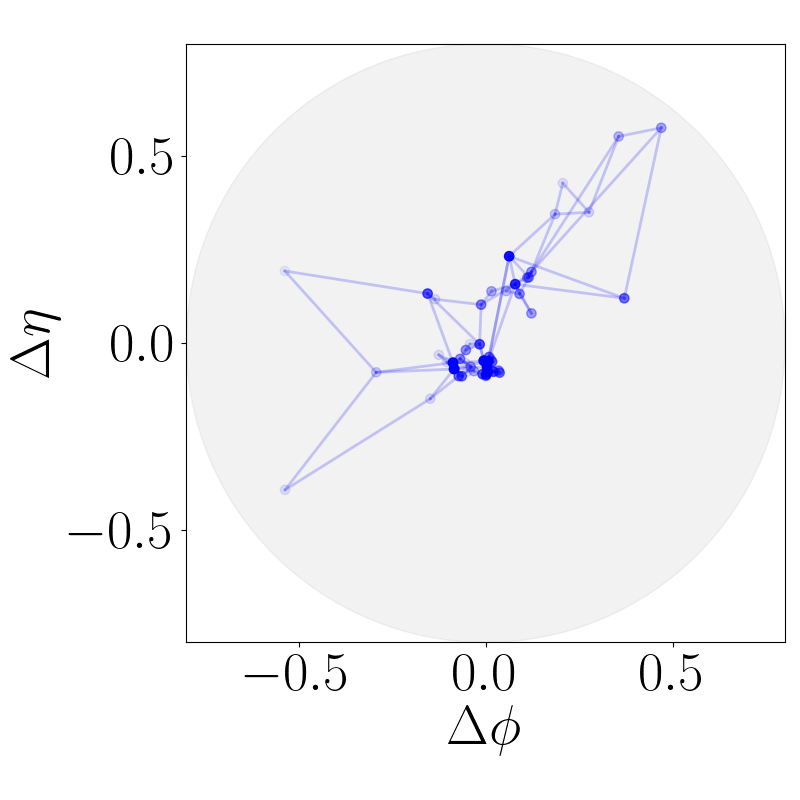}
\includegraphics[width=0.32\textwidth]{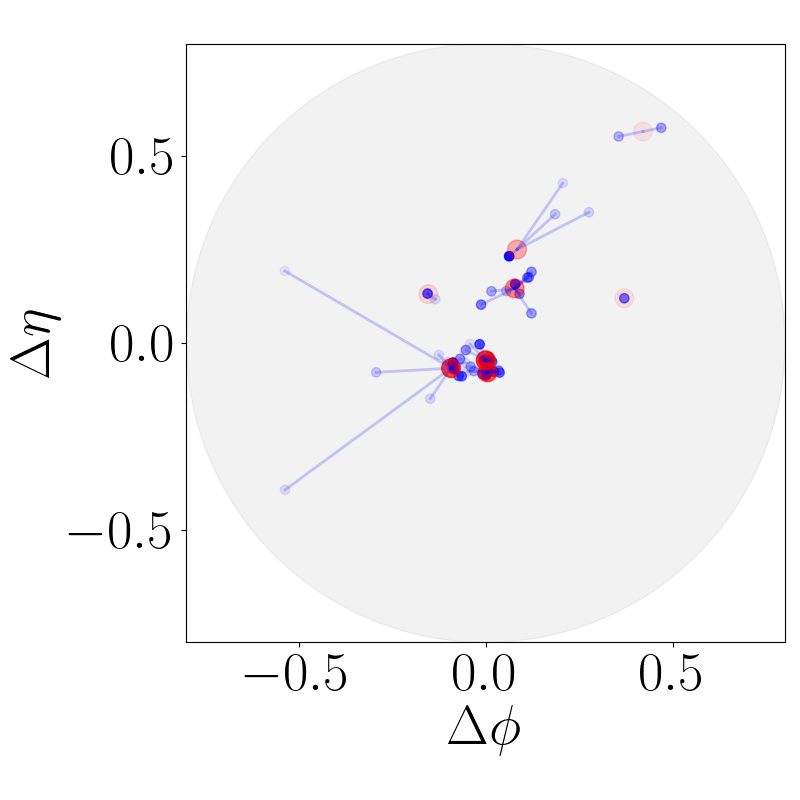}
\includegraphics[width=0.32\textwidth]{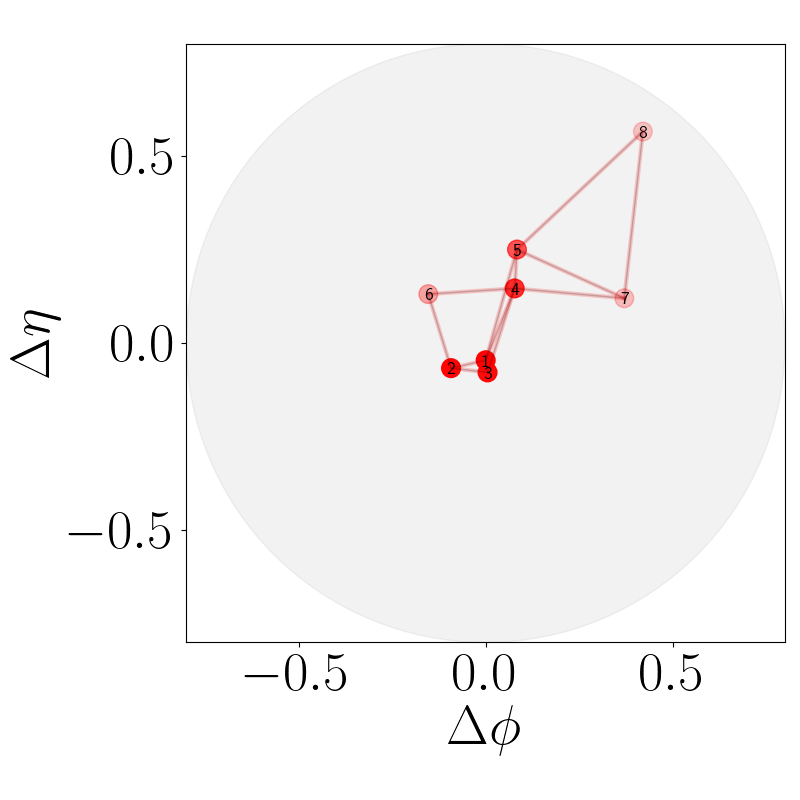}
\caption{Representation of a two-pronged $Z$ boson jet as a graph. Left panel: hadrons (constituents of a jet) shown as a Laman graph built using the Henneberg construction, described in section \ref{sec:graphconstruction} and Fig.~\ref{fig:laman_proximity}. Middle panel: hadrons grouped into 8 exclusive $k_{T}$ subjets, which are shown as red nodes. The blue edges connect hadrons to the subjet they belong to. Right panel: the Laman graph built from subjets according to the same procedure as the left panel. The intensity of particle and subjet nodes is proportional to the fraction of the jet $p_{T}$ they carry. The coordinates are in plane defined by the rapidity $\eta$ and the azimuthal angle $\varphi$ with respect to the jet axis where $\Delta \eta = \eta^{\mathrm{node}} - \eta^{\mathrm{jet}}$ and $\Delta \phi = \phi^{\mathrm{node}} - \phi^{\mathrm{jet}}$.}
\label{fig:subjetsandgraphs}
\end{figure*}

In this work, we develop an unsupervised anomaly detection algorithm based on a graph autoencoder. Rather than reconstructing the absolute three-momenta of particles or subjets, our model focuses on their transverse momenta $p_T$ and relative angular distances in the rapidity–azimuth plane. Using relative distances (either between particles or with respect to the jet axis) rather than absolute positions is more closely related to observables calculable in perturbative QCD. Moreover, high-energy jet structure is expected to be encoded primarily in relative features rather than absolute ones, which we confirm by additionally training classifiers on several representative jet physics tasks.

The primary focus of our work is on the use of different graph representations of jets. Instead of relying only on fully connected graphs that encode all pairwise distances, we explore the use of sparse graphs for anomaly detection algorithms. For the sparse graphs we consider, the number of edges scales linearly with the number of particles or subjets, whereas for fully connected graphs it scales quadratically. Specifically, we consider so-called locally rigid Laman graphs and different globally rigid unique graphs. The notion of rigidity is related to the embedding of a graph in Euclidean space, here specifically $\mathbb{R}^2$, given only relative distances between nodes. The two types of sparse graphs we consider correspond to the minimal sets of constraints required to determine such an embedding, as discussed in more detail below. Due to their connection to the minimal amount of information needed to specify the graph structure in $\mathbb{R}^2$ or the azimuth-rapidity plane, they provide a natural choice for a physically motivated inductive bias in graph-based autoencoders for anomaly detection. Similar to the clustering of jet constituents into subjets, these graph constructions allow us to control the amount of information accessible to the machine while ensuring that sufficient geometric structure is retained. We find that certain unique graph constructions achieve the best performance, as measured by the maximum Significance Improvement Characteristic (SIC) curve on the LHC Olympics data set, which we use to benchmark our approach.

The remainder of this work is organized as follows. In section~\ref{section:GraphTheory}, we introduce the representation of jets in terms of graphs based on particles or reclustered subjets. In addition, we review relevant concepts from graph theory, such as graph rigidity and the graph embedding problem. In particular, we describe both Laman and different unique graph construction techniques that will be used as input to the machine learning-based anomaly detection algorithm. In section~\ref{section:AnomalyDetection}, we describe the graph autoencoder developed in this work for unsupervised anomaly detection and present numerical results evaluating its performance on the LHC Olympics data set. We conclude in section~\ref{section:conclusions} and provide an outlook. In Appendix~\ref{section:appendix}, we additionally present results on jet classification using the different graph construction techniques.

\section{Graph theory and jet representations}
\label{section:GraphTheory}

We start by discussing the representation of jets that will serve as input to the machine-learning algorithms described in the next section. First, we cluster jets into subjets to control the amount and the complexity of information that the machine has access to. Next, we will introduce relevant concepts in graph theory and describe different graph construction techniques that will be used as the basis of the graph autoencoder developed in the subsequent section.

\subsection{Subjet clustering and phase space considerations}

The set of massless particles contained in a high-energy jet can be described in terms of their transverse momenta, absolute positions in rapidity, and azimuthal angle $(p_{Ti},\eta_i,\phi_i)$. Due to rotational and translational invariance, we can also represent the particles inside a jet in terms of their transverse momenta $p_{Ti}$, and the relative distances between all particles $i$ and $j$ in the rapidity-azimuth plane $\theta_{ij}=(\Delta\eta_{ij}^2+\Delta\phi_{ij}^2)^{1/2}$. Jet clustering algorithms are designed to recursively cluster soft and collinear emissions to approximately reconstruct the parton branching mechanism that led to the observed jet. To systematically throttle the information used as input to machine learning algorithms, we recluster particles into a given number of subjets. We cluster particles using the exclusive $k_T$ algorithm with the $E$ recombination scheme~\cite{Catani:1993hr}. Here, the recursive clustering algorithm is stopped when a given number of subjets is obtained. This results in a fixed number of subjets per jet but with varying subjet radii. Subjets have been explored in the context of jet classification tasks to assess the relevance of Infrared-Collinear (IRC) safe information in Ref.~\cite{Athanasakos:2023fhq}. In this work, we will use subjet reclustering to throttle the information that the anomaly detection algorithm has access to, as described in the next section. By changing the number of reclustered subjets, we can systematically dial in the amount of soft and collinear information that the machine learning algorithm has access to. This approach allows for a smooth interpolation to the hadron level where the full information content is accessible. The use of IRC safe subjets is closely related to the use of complete sets of jet substructure observables such as the $N$-subjettiness~\cite{Thaler:2010tr,Stewart:2010tn} basis~\cite{Datta:2017rhs,Datta:2017lxt,Datta:2019ndh} and Energy Flow Polynomials~\cite{Komiske:2017aww}. See Fig.~\ref{fig:subjetsandgraphs} for an illustration of the particle content of a representative high-energy jet at the LHC in the rapidity-azimuth plane. The middle panel shows the assignment of hadrons grouped into 8 exclusive $k_T$ subjets. The graph structures shown in the left and right panel of Fig.~\ref{fig:subjetsandgraphs} will be discussed in more detail below.

In general, a machine learning algorithm for classification or anomaly detection tasks can take as input the four momenta of all $N$ particles inside the jet, which corresponds to $4N$ variables per jet. For massless particles using momentum conservation, this is reduced to $3N-4$ independent variables. This $(3N-4)$-dimensional phase space manifold was found to be isomorphic to $\Delta_{N-1} \times S^{2N-3}$~\cite{Cai:2024xnt, Larkoski:2020thc}. Here, $\Delta_{N-1}$ corresponds to an $N-1$ dimensional simplex, which means that the $N$ transverse momenta add to the total jet momentum, and $S^{2N-3}$ denotes the $2N-3$ dimensional sphere. This manifold can be parametrized in terms of $N-1$ transverse momenta $p_{Ti}$, and a suitable basis of $2N-3$ relative angles, $\theta_{ij}$, between particle pairs~\cite{Datta:2017rhs}. This manifold is related to so-called Laman graphs but not unique graphs, as discussed in the next section.

In principle, a machine learning algorithm has access to the same information content using the absolute positions encoded in the full particle four momenta or in terms of transverse momenta and relative distances. However, in practice, the performance of machine learning algorithms for various tasks at collider experiments can differ significantly depending on the type and amount of input information that is provided and the architecture using the provided information. For example, machine learning-based jet classification algorithms typically achieve the best performance the more low-level information is provided~\cite{Datta:2017rhs,Lai:2021ckt}. However, it has been found that unsupervised anomaly detection algorithms at collider experiments can perform well with a limited number of high-level observables. See for example Ref.~\cite{Stein:2020rou}. In this work, instead of hand selecting a set of high-level observables, we will employ reclustered subjets, which allows us to smoothly interpolate between high-level information analogous to observables or summary statistics and low-level particle information. We find that the performance of our graph autoencoder for anomaly detection peaks at intermediate values of the provided information. In addition, from our numerical studies, we find that we achieve a better performance by providing only relative distances of particles instead of their absolute positions. 

\subsection{Graph rigidity, Laman and unique graphs}
\label{sec:phasespace_rigidity} 

Unsupervised anomaly detection at collider experiments may benefit from enforcing physically motivated inductive biases in the machine learning architecture. We will explore these concepts numerically using a graph neural network (GNN)-based autoencoder in the next section. First, since tasks like jet classification and anomaly detection at collider experiments are invariant under translations and/or rotations, it is natural to work with relative distances between particles or subjets, rather than absolute positions in the rapidity–azimuth plane as discussed above. When representing jets as graphs, the relative distances are naturally encoded in the edge features of the graph. Second, the choice between fully connected and sparse graphs impacts the expressiveness and generalization of the downstream machine learning architecture. For anomaly detection, we aim to achieve a balance: the graph should be sparse enough to avoid redundancies, yet sufficiently connected to preserve essential geometric structures. These considerations motivate a graph-theoretic framework for jet representations. To formalize this, we are going to review several fundamental concepts from graph theory, which will motivate the construction of certain sparse graph structures, which we will explore in the context of anomaly detection in jet physics.

\begin{figure*}[t]
\centering
\includegraphics[width=0.55 \textwidth]{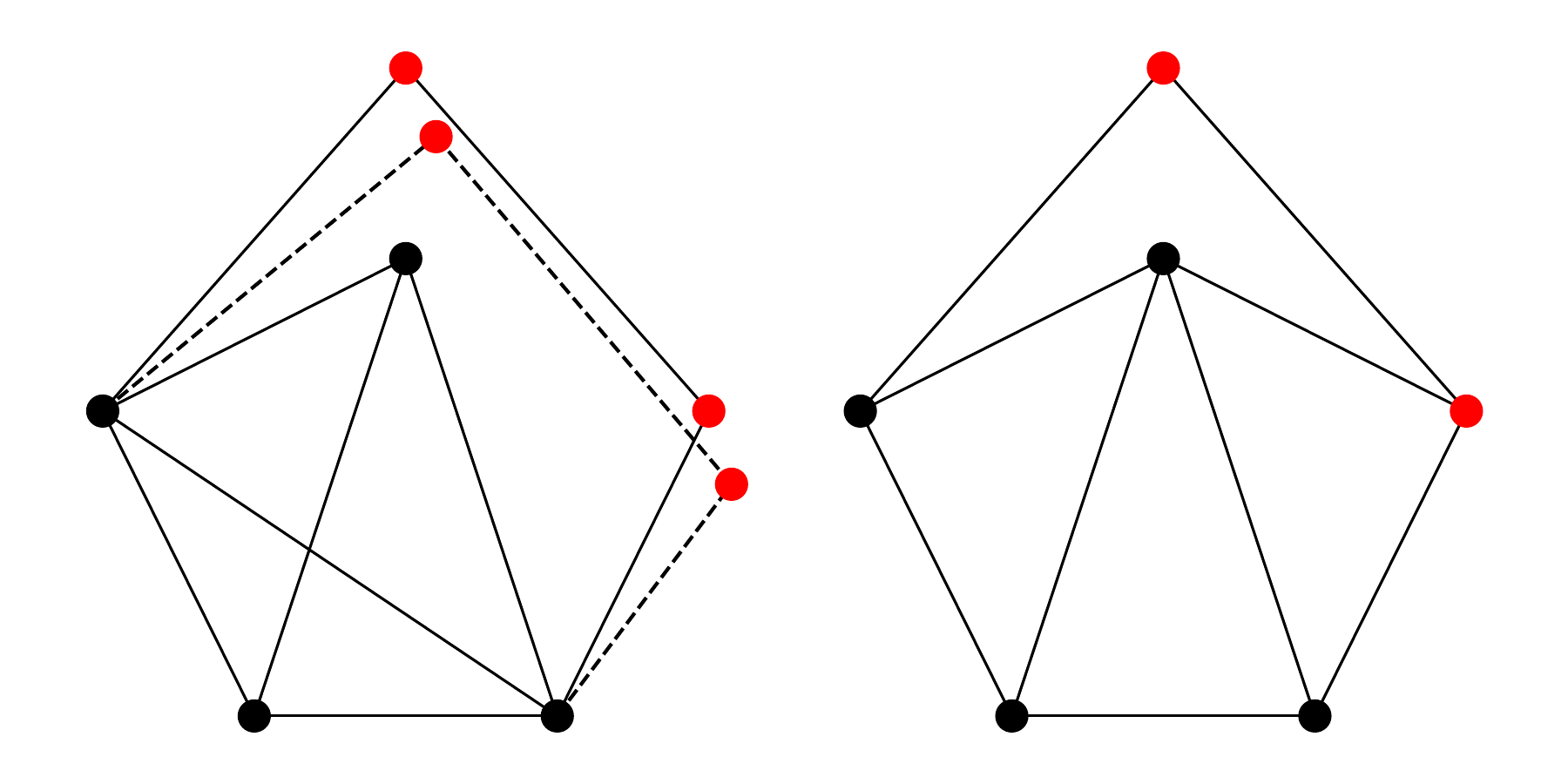} 
\caption{Representative examples of flexible (left) and locally rigid (right) graphs with $N=6$ nodes and $|E|=2N-3=9$ edges embedded in $\mathbb{R}^2$. The flexible graph admits continuous deformations of the two red vertices that preserve pairwise edge lengths while changing distances between non-adjacent nodes. Two distinct embeddings of the same flexible graph are shown (solid and dashed lines). In contrast, by moving one of the edges, the right graph is locally rigid.}
\label{fig:flexible_twoedges}
\end{figure*}

Let $G=(E,V)$ denote an undirected graph, where $V$ is a set of $N=|V|$ vertices, and $E$ is a set of $m=|E|$ edges, encoding pairwise geometric relations. Each particle or subjet $i\in V$ is associated with scalar features, such as the normalized transverse momentum $p_{Ti}$, and each edge $(i,j)\in E$ corresponds to a measured relative distance or angle of particles $\theta_{ij}$ and/or similar relational quantities. There are various ways to choose the connectivity structure of the graph. For example, choosing only a specific subset of $2N-3$ edges yields a graph that minimally captures geometric constraints. Alternatively, a fully connected graph includes all ${N \choose 2}=N(N-1)/2$ pairwise distances, as used in transformer-based models. While fully connected graphs are more expressive, we find that anomaly detection performance can improve when using certain sparse graphs. To motivate how to choose such graphs, we are now going to review the concept of rigid vs. flexible graphs, which is tied to how graphs can be embedded in Euclidean space.

\begin{figure*}[t]
\centering
\includegraphics[width=0.7 \textwidth]{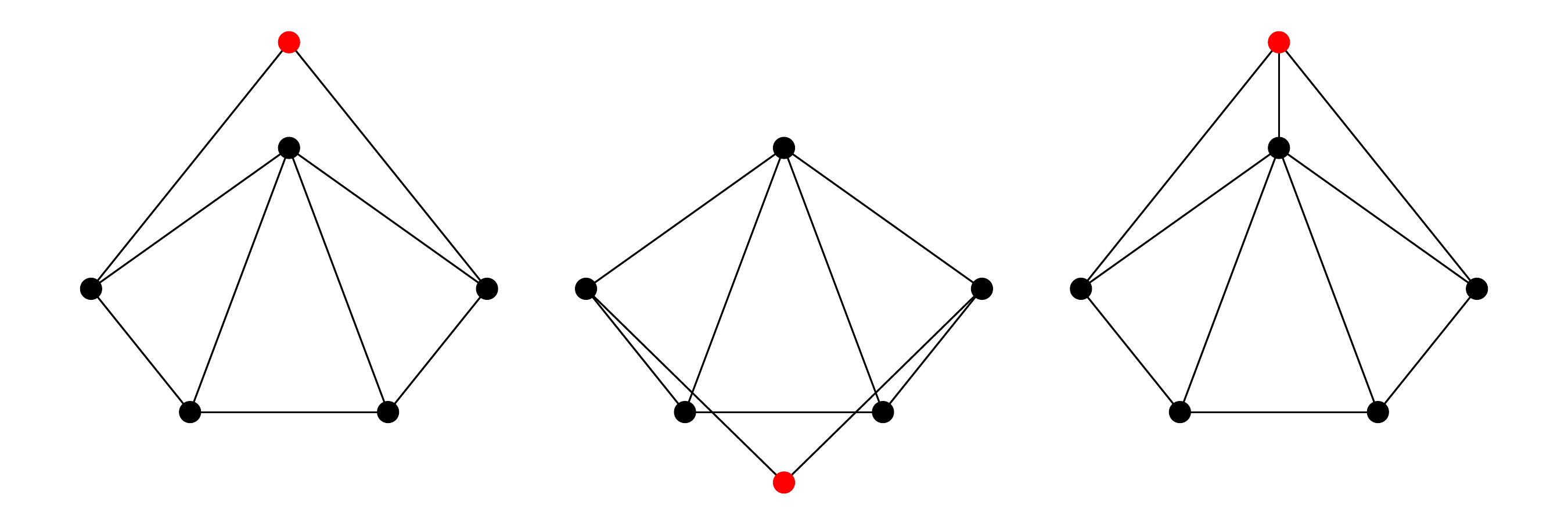}
\caption{Illustration of locally vs. globally rigid graphs embedded in $\mathbb{R}^2$. The two graphs on the left correspond to two distinct realizations of the same locally rigid Laman graph. Reflecting the red node about the horizontal axis preserves all pairwise edge lengths but alters the distances between non-adjacent nodes. Instead, the graph on the right is globally rigid, which allows for only one unique realization up to the isometries.~\label{fig:rigidvsglobal}}
\end{figure*}

A central question in graph theory and graph reconstruction is whether a graph with given edge lengths can be embedded in an Euclidean space. Formally, given 
\begin{itemize}
    \item an undirected graph $G=(V,E)$,
    \item a set of edge lengths $\{e_{ij}\}$ with $(i,j)\in E$,
    \item the dimension $d$ of the Euclidean space $\mathbb{R}^d$,
\end{itemize}
find a mapping $p:V\to\mathbb{R}^d$ such that
\begin{equation}
    ||p(i)-p(j)||=e_{ij},\;\; \forall (i,j)\in E\,.
\end{equation}
Here $p(i)$ denotes the position of node $i$ and $||\cdot||$ denotes the Euclidean distance metric in $\mathbb{R}^d$. That is, we are looking for the positions of nodes that are consistent with the given edge lengths. In the context of jet physics, this corresponds to embedding particles or subjets in the rapidity-azimuth plane, such that their given relative distances or angles are preserved. A key question is whether a given graph, with its set of pairwise distances, admits a unique realization (up to global isometries such as translations and rotations), or whether multiple, potentially infinitely many, configurations exist that satisfy the same set of edge constraints. This is described by the notion of graph rigidity. A graph is called
\begin{itemize}
    \item {\it Flexible} if it can be continuously deformed while preserving pairwise edge constraints, resulting in an infinite family of realizations. This flexibility corresponds to degrees of freedom in the relative configuration of non-adjacent particles that are not connected by edges. For an illustration, see Fig.~\ref{fig:flexible_twoedges}.
    \item {\it Locally rigid} if there are finitely many discrete realizations. See the left and middle graph in Fig.~\ref{fig:rigidvsglobal}, where the reflection of one node leads to two distinct realizations of a graph embedded in $\mathbb{R}^2$. Rigid graphs are locally unique: small perturbations of the vertex positions that preserve the edge lengths result only in isometric configurations.
    \item {\it Globally rigid} if there is only one unique embedding up to isometries. For an illustrative example comparing locally and globally rigid graphs, see Fig.~\ref{fig:rigidvsglobal}.
\end{itemize}

In the context of both jet classification and anomaly detection, we expect that machine learning algorithms based on flexible graphs generally underperform since they do not contain sufficient geometric information. Instead, we will focus on rigid graphs in this work. Since our application lies in two spatial dimensions, embedding a jet in the rapidity–azimuth plane, we now restrict our discussion to graph embeddings in $\mathbb{R}^2$. In this case, the minimal number of edges required to achieve local rigidity is $|E|=2N-3$. Minimally rigid graphs are referred to as {\it Laman graphs} \cite{Laman}. Minimal rigidity means that removing one edge always results in a flexible graph. Since Laman graphs ensure local rigidity, they are a natural candidate for a machine-learning architecture based on sparse graphs.

While Laman graphs ensure local rigidity, they may still admit multiple discrete realizations in $\mathbb{R}^2$. Global rigidity is obtained for {\it unique graphs}. A graph has a unique realization in $\mathbb{R}^2$ if and only if it satisfies Hendrickson's conditions \cite{Hendrickson, Connelly2005, gortler2010characterizing}. First, the graph must be redundantly rigid, meaning it remains rigid after the removal of any single edge. Second, it must be 3-connected, i.e. for every pair of nodes there exist three vertex-disjoint paths between them. As a result, unique graphs may be ideally suited for machine learning tasks at collider experiments as they encode enough relational geometry to fix all positions unambiguously up to isometries.

In contrast to fully connected graphs, finding an embedding of a sparse graph in $\mathbb{R}^2$ that satisfies given edge length constraints is, in general, an NP-hard problem in the worst case~\cite{saxe1979embeddability, Hendrickson, inproceedings}. Designing a machine learning architecture for anomaly detection that is tasked to solve this embedding problem could provide new insights into the performance of the corresponding algorithm. However, for practical applications, the relevant computational complexity is determined more by the average case than by worst-case hardness. A detailed investigation of this aspect is beyond the scope of this work. Instead, we will focus on anomaly detection algorithms based on Laman and unique graph constructions as discussed in the next section.

\subsection{Graph constructions for jets}
\label{sec:graphconstruction}

In this section, we discuss in detail graph construction techniques for jets, focusing on Laman and unique graphs. When constructing graphs, we generally rely on an ordering of the particles or subjets. Throughout this section, we order particles/subjets by their transverse momentum $p_T$, starting with the particles with the highest $p_T$. We also explored alternative orderings but consistently observed worse performance for the machine learning tasks described below. Note that this ordering breaks permutation invariance for the sparse graphs considered here. Permutation invariance is only recovered in the limit of fully connected graphs.

We start by discussing the construction of Laman graphs, which are minimally rigid graphs with $2N-3$ edges. All Laman graphs can be obtained by the Henneberg construction, which provides an inductive recipe using two types of operations~\cite{henneberg1911graphische, HAAS200531}. Starting from a single edge between two nodes:

\begin{itemize}
\item Type I: Add a new node and connect it to two existing nodes with new edges.
\item Type II: Subdivide an existing edge, then connect the new node to a third existing node.
\end{itemize}

\begin{figure}[t]
    \centering
    \begin{tabular}{c c}
        \begin{minipage}{0.4\textwidth}
            \includegraphics[width=\textwidth]{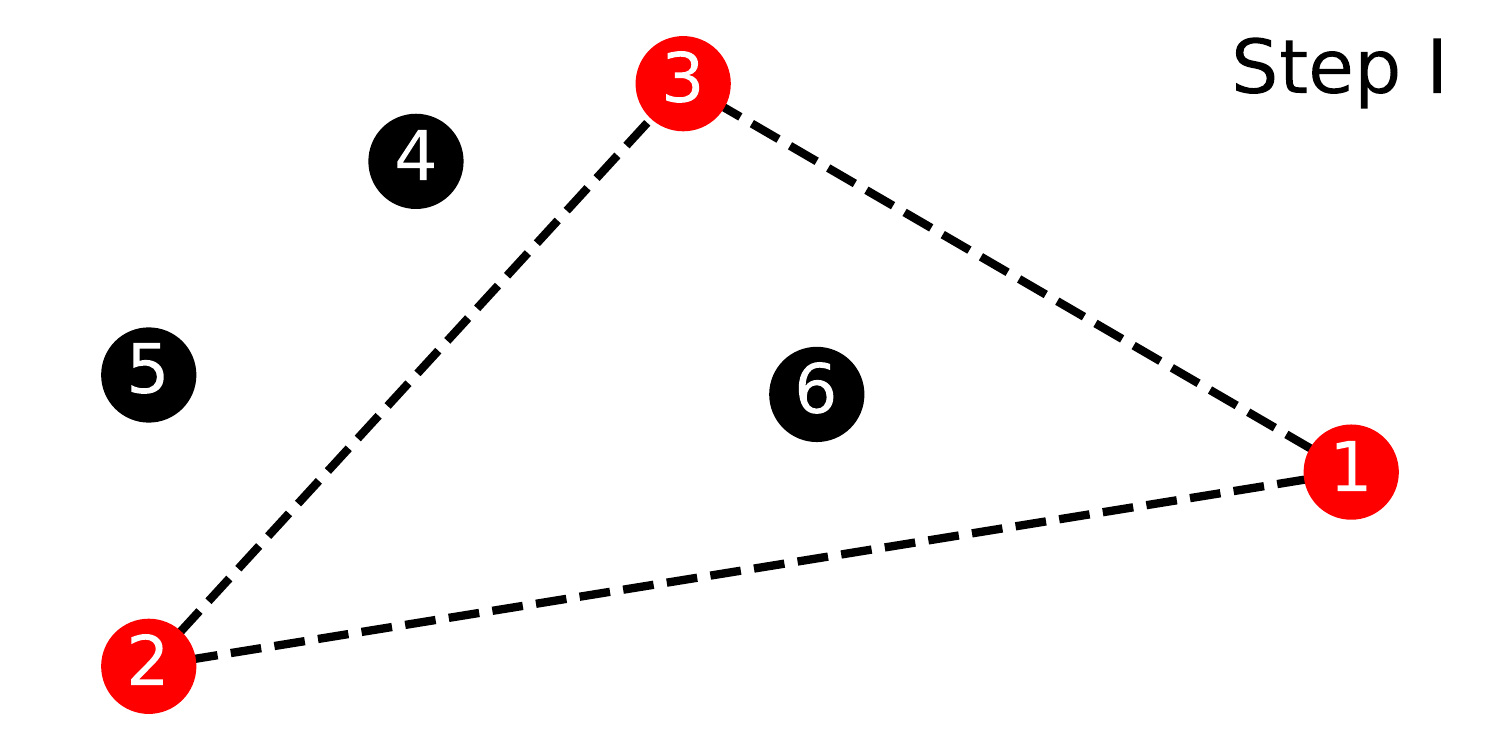}
        \end{minipage} &
        \hspace*{1cm}
        \begin{minipage}{0.4\textwidth}
            \includegraphics[width=\textwidth]{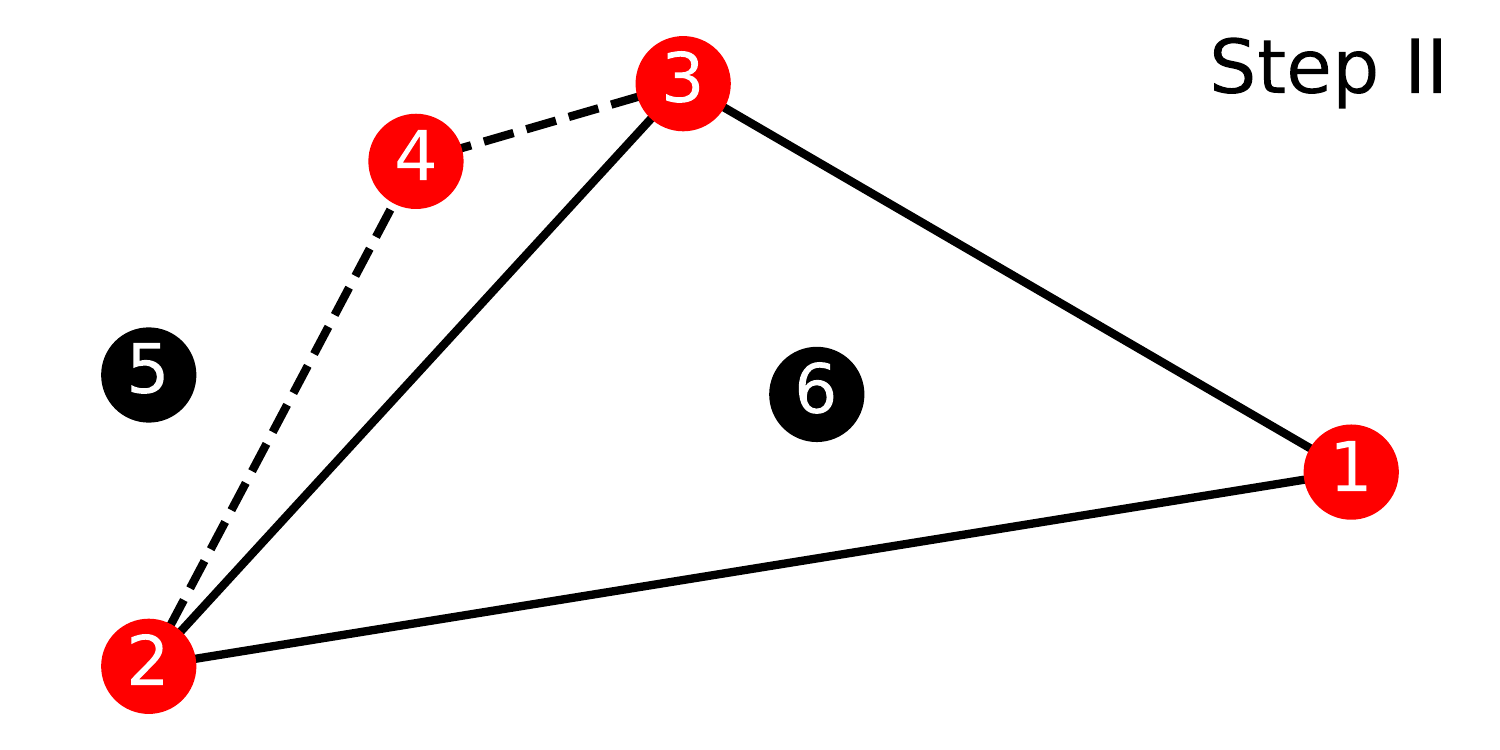}
        \end{minipage} \vspace{1cm} \\
        \begin{minipage}{0.4\textwidth}
            \includegraphics[width=\textwidth]{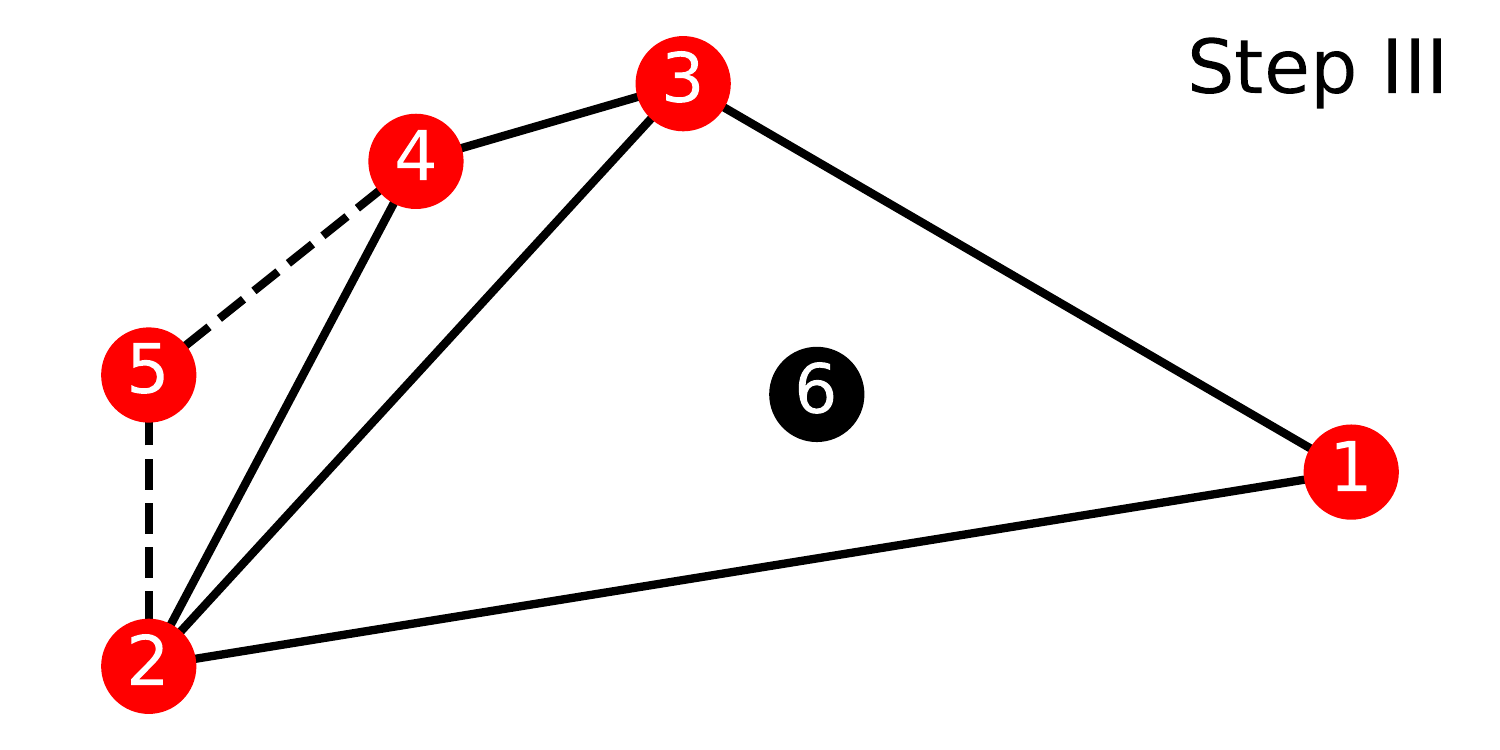}
        \end{minipage} &
        \begin{minipage}{0.4\textwidth}
            \includegraphics[width=\textwidth]{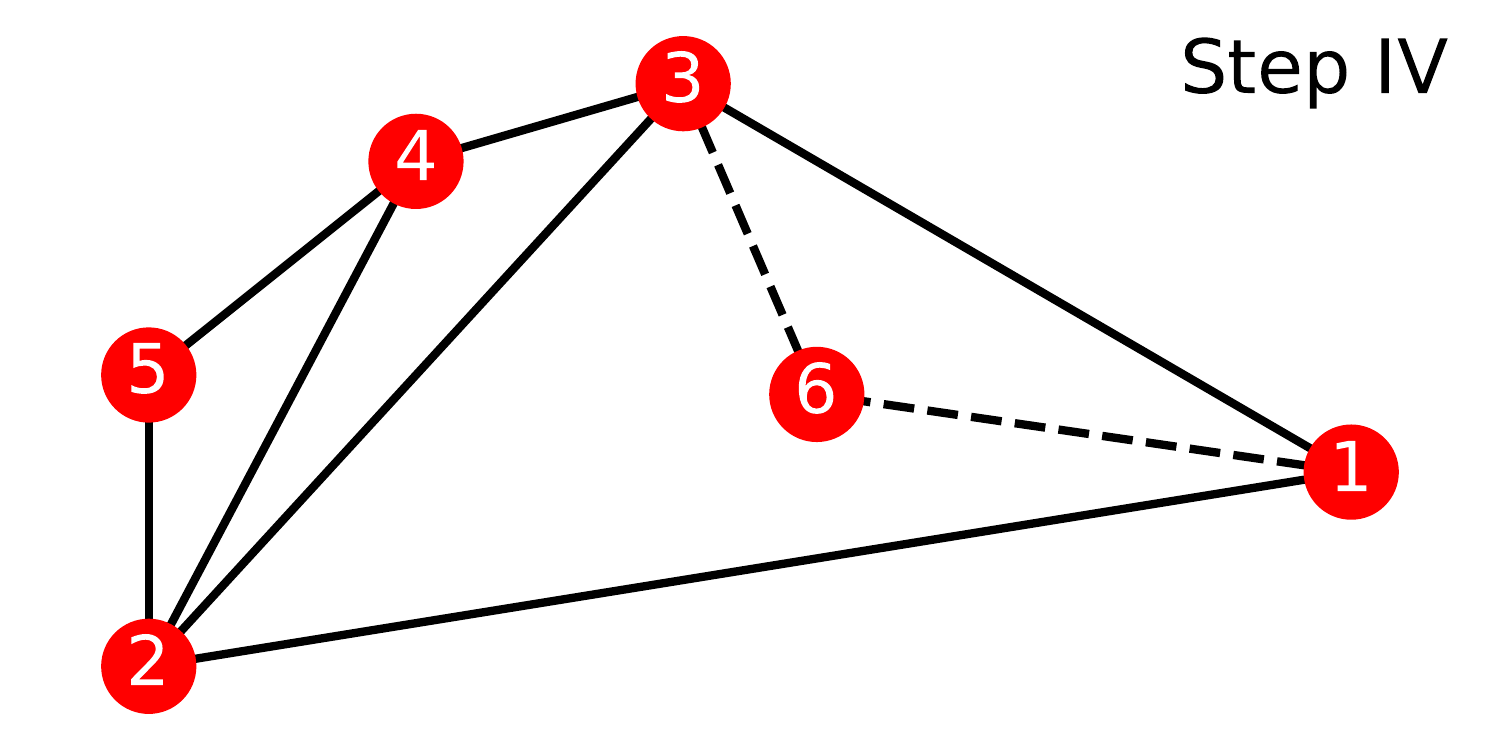}
        \end{minipage} \\
    \end{tabular}
    \caption{Illustration of the Laman graph construction for a jet with six particles (nodes), ordered by descending transverse momentum $p_{T}$. The three hardest particles form an initial triangle. Each subsequent particle is added to the graph by connecting it to its two nearest neighbors, based on angular proximity in the $\eta$-$\phi$ plane, among the previously included nodes. This sequential procedure ensures that the resulting graph is locally rigid, with $2N - 3$ edges.} 
    \label{fig:laman_proximity}
\end{figure}

For simplicity, we restrict ourselves to Type I operations, which are sufficient to construct Laman graphs. See Fig.~\ref{fig:laman_proximity} for an illustration. We begin with a triangle formed by the three hardest particles, i.e., those with the highest $p_T$. We then iterate through the list of remaining particles, adding them one by one to the existing graph. Each new particle is connected to the graph with two edges linking it to existing nodes. There are multiple ways to choose the two existing nodes to connect to each new particle. For instance, one could always connect to the two hardest particles. However, we find improved performance for the machine learning task discussed below when each new particle is connected to its two nearest neighbors in angular distance in the $\eta$–$\phi$ plane. This procedure is similar to the Cambridge/Aachen jet clustering algorithm~\cite{Dokshitzer:1997in,Wobisch:1998wt}. Using a Voronoi tesselation, this can be achieved in $\mathcal{O}(N \ln N)$ steps. See Fig.~\ref{fig:subjetsandgraphs} for a representative Laman graph using the Henneberg construction for the particles (left panel) and subjets (right panel) inside a jet. The resulting graph has $2N-3$ edges and is locally rigid.

\begin{figure}[t]
    \centering
    \begin{tabular}{ c c }
        
        \begin{minipage}{0.4\textwidth}
            \includegraphics[width=\textwidth]{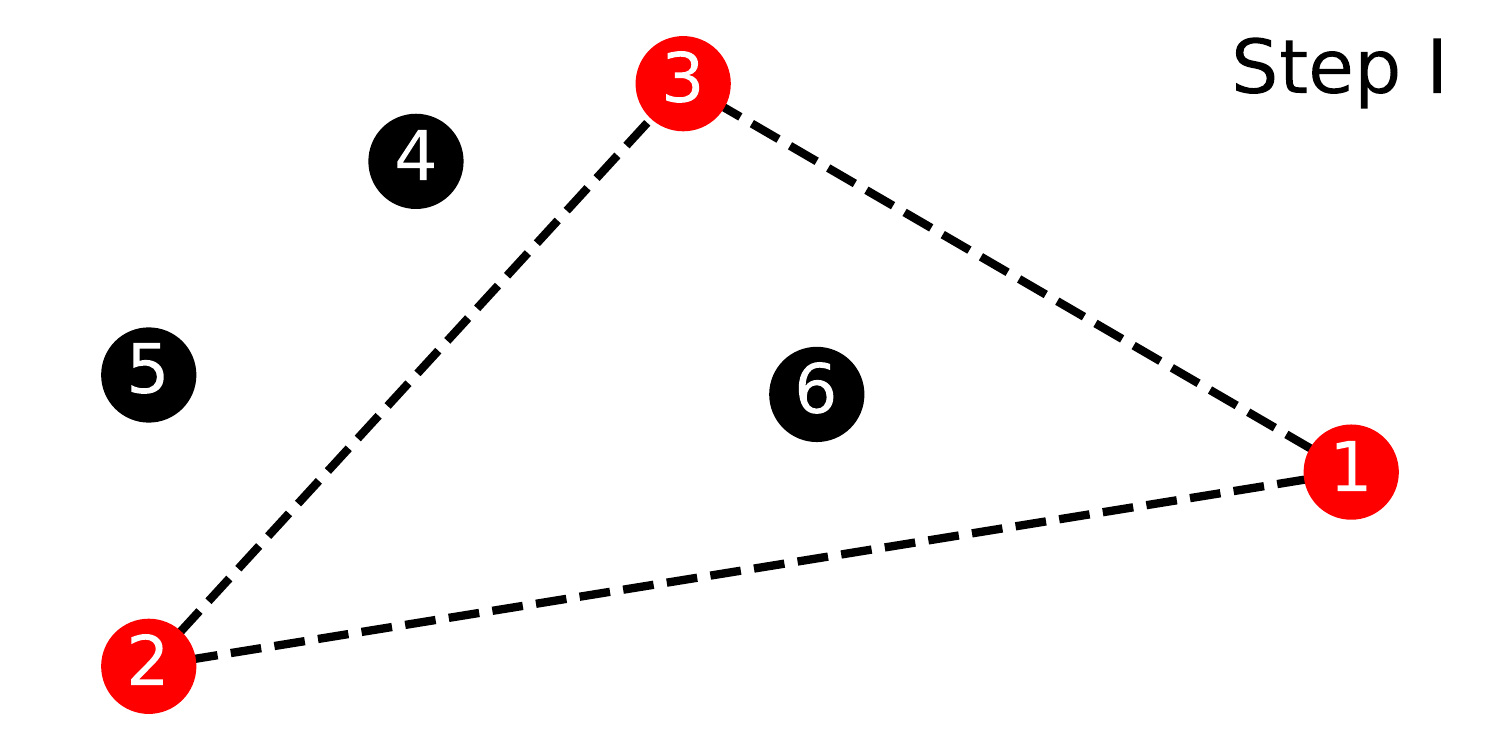}
        \end{minipage} &
        \hspace*{1cm}
        \begin{minipage}{0.4\textwidth}
            \includegraphics[width=\textwidth]{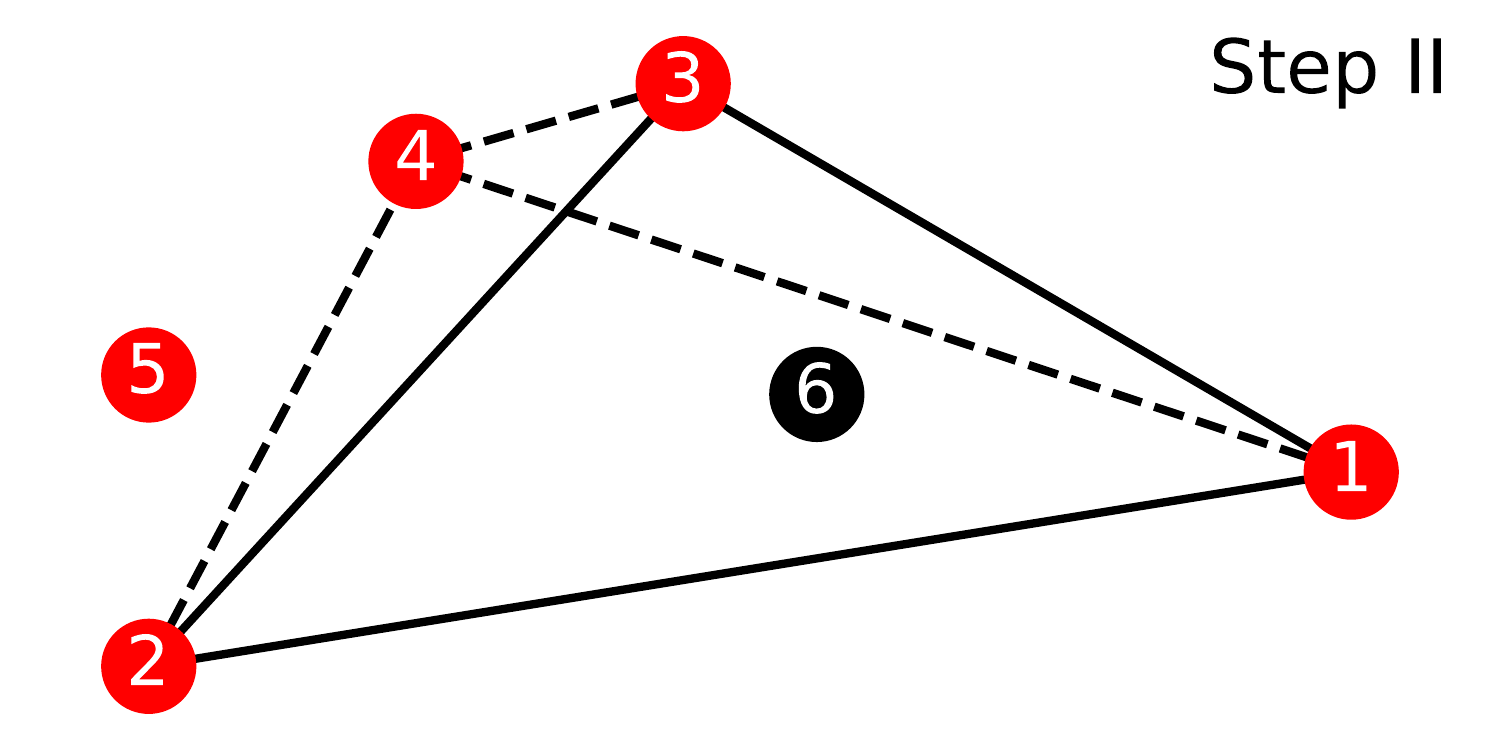}
        \end{minipage} \vspace{1cm} \\
        \begin{minipage}{0.4\textwidth}
            \includegraphics[width=\textwidth]{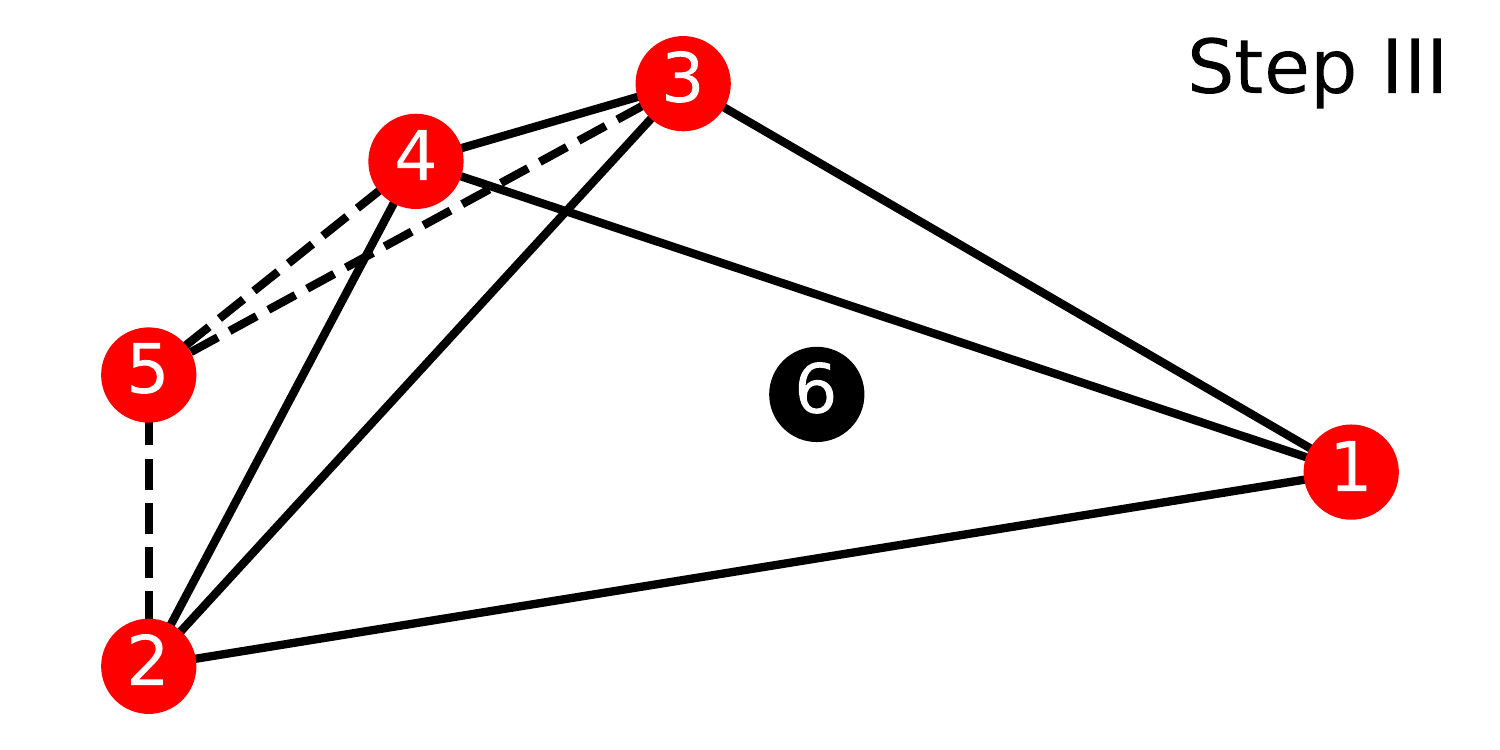}
        \end{minipage} &
        \begin{minipage}{0.4\textwidth}
            \includegraphics[width=\textwidth]{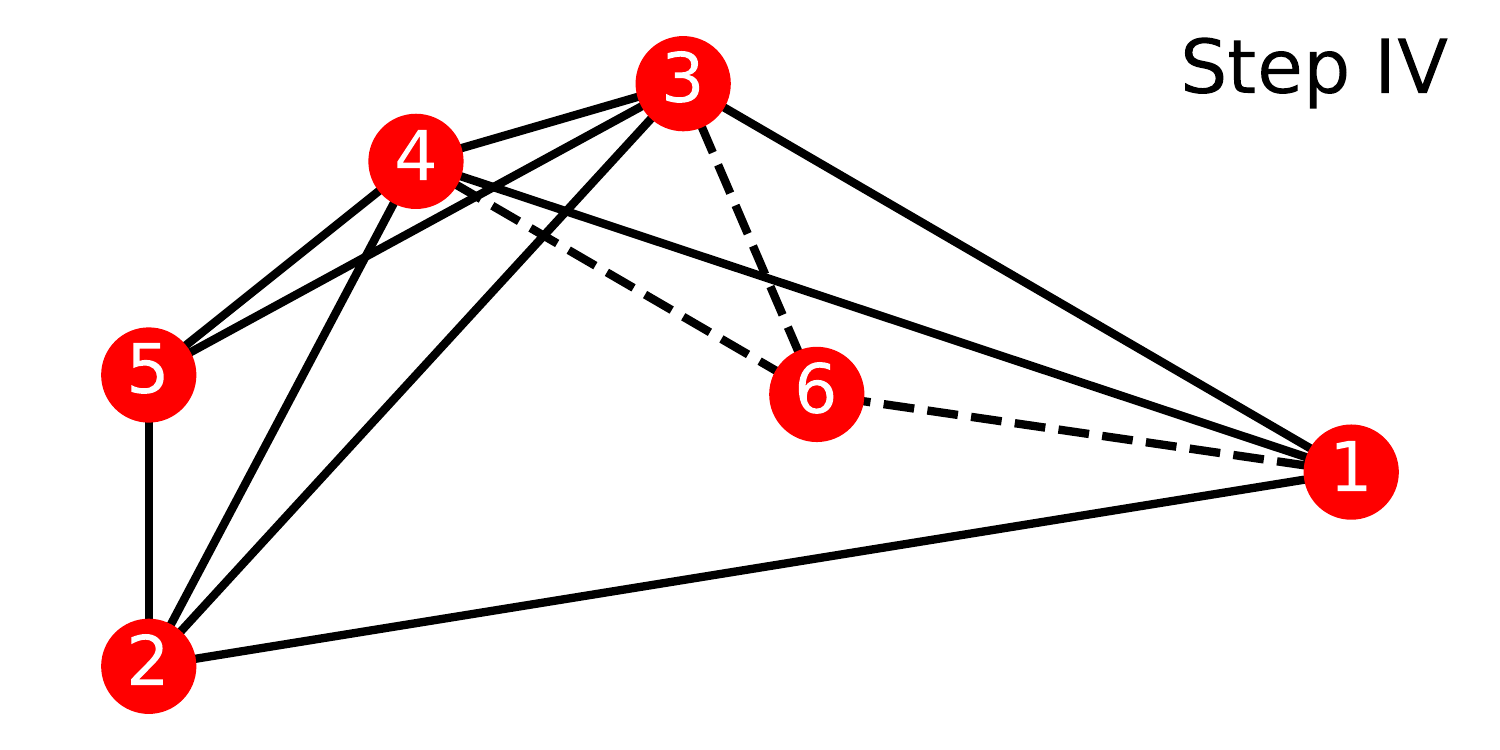}
        \end{minipage} \\
    \end{tabular}
    \caption{Construction of a \textit{unique-3} graph analogous to Fig.~\ref{fig:laman_proximity}. The resulting graph is globally rigid allowing for a unique realization in $\mathbb{R}^2$ up to isometries.} 
    \label{fig:un3_proximity}
\end{figure}

Analogously, we can construct unique graphs that are globally rigid with $|E|={\cal O}(3N)$ edges. We again sort particles by decreasing $p_T$, and start with a fully connected graph of the four hardest particles. Each subsequent particle is then connected to its three nearest neighbors in angular distance, adding three edges per new node. Repeating this process yields a graph that satisfies the Hendrickson conditions for global rigidity. See Ref.~\cite{Hendrickson} for more details. In the following, we refer to this construction as a {\it unique-3} graph. More generally, we define \textit{unique-k} graphs by starting from a fully connected subgraph of the $k+1$ hardest particles. For each new particle, we add $k$ edges to its $k$ nearest neighbors. The resulting graph has $|E|=kN-k(k+1)/2$ edges. See Fig.~\ref{fig:un3_proximity} for an illustration of the {\it unique-3} graph construction. For $k\geq 3$ the resulting graph is globally rigid, whereas for $k=2$, we recover the locally rigid (not unique) Laman graph. Note that these unique graphs are not minimal constructions of globally rigid graphs, which are referred to as $M$-circuits~\cite{JACKSON20051}. They can be obtained using a modification of the Type II Henneberg operations. However, in this work, our primary concern is whether the graph admits a unique realization, rather than its minimality. While increasing $k$ yields denser graphs, the number of edges still scales linearly with $N$, in contrast to the ${\cal O}(N^2)$ scaling of fully connected graphs. In the next section, we are going to use the graph constructions introduced here for anomaly detection tasks at the LHC using a graph-based autoencoder.

\section{Anomaly detection at the LHC}
\label{section:AnomalyDetection}

We begin by briefly reviewing the LHC Olympics data set~\cite{lhco, kasieczka_2019_4536624}, which serves as the benchmark for our numerical studies in the subsequent sections. Next, we describe the machine learning architecture of the graph autoencoder used throughout this work. We then present the results of our anomaly detection analysis using different subjet clusterings and graph constructions.

\subsection{Benchmark LHC anomaly detection data set}
\label{sec:anomaly_dataset}

As the benchmark data set for anomaly detection studies of BSM signals at the LHC, we use the LHC Olympics challenge 2020 data set~\cite{lhco, kasieczka_2019_4536624}, which enables direct comparison with a wide range of recent methods~\cite{Andreassen:2020nkr, cathode, Buhmann:2023acn, Hallin:2022eoq, Stein:2020rou, cheng2025incorporatingphysicalpriorsweaklysupervised, Golling:2023yjq, Golling:2022nkl, dAgnolo:2021aun, Nachman_2020, Andreassen_2020, metodiev2024anomalydetectioncolliderphysics, Amram:2020ykb}. It contains up to 1M background events generated from Standard Model processes, resulting in dijet final states. The anomalous signal events are created using the resonant production of a new heavy boson $Z'$, that decays as $Z' \to X(\to q \bar{q}), Y(\to q \bar{q})$, with masses $m_{Z'} = 3.5$~TeV, $m_X = 0.5$~TeV, and $m_Y = 0.1$~TeV. 

\begin{figure*}[t]
\centering
\includegraphics[width=0.8 \textwidth]{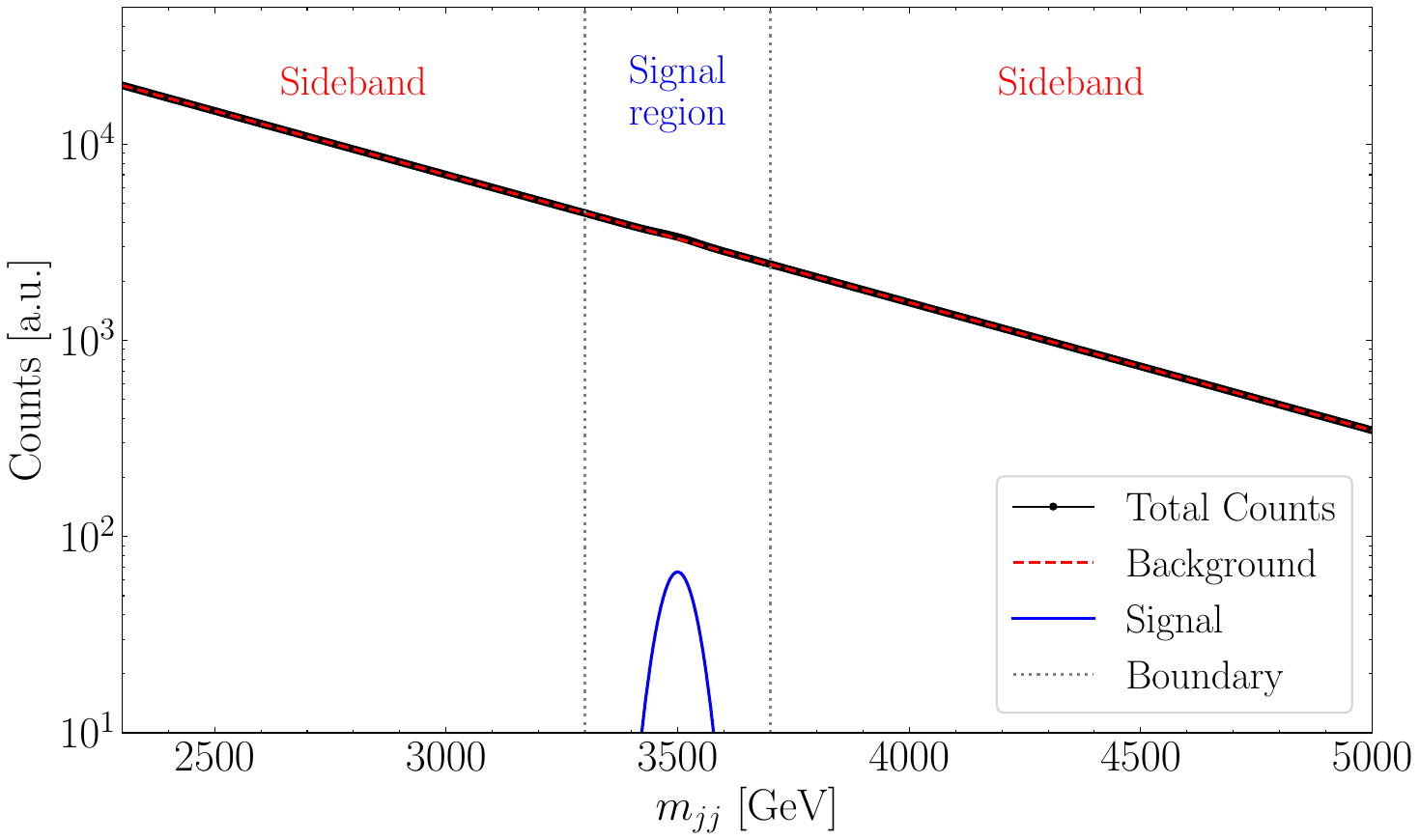}
\caption{Schematic depiction of the dijet mass distribution of background (red) and signal (blue) in the LHC Olympics data set. The sideband regions are dominated by background events, while the signal region contains a finite fraction of signal events. Here, it corresponds to $S/B\sim 3\cdot10^{-3}$ in the signal region region and a significance $S/\sqrt{B} \sim \mathcal{O}(1)$.} 
\label{fig:mass_dist}
\end{figure*}

The BSM particle $Z'$ is produced on-shell and decays into two intermediate particles $X,Y$ that subsequently decay into quarks. This produces a signal characterized by a peak in the dijet invariant mass spectrum. For sufficiently large signal fractions, the $Z'$ resonance can be discovered using traditional bump-hunting techniques. Figure~\ref{fig:mass_dist} provides an illustrative representation of a bump-hunt search. The signal region is identified as the area with the highest concentration of signal events, while the sidebands serve as control regions where the signal contribution is negligible. The blue, black, and dashed red lines represent the signal, observed counts, and background yields, respectively. Dashed vertical lines delineate the signal and sideband regions. For smaller signal fractions, the bump’s significance is reduced and detection may instead require machine learning techniques that can exploit the full event information. This type of data set naturally fits within the weakly-supervised anomaly detection paradigm, where the sideband regions are dominated by background events, while the signal region is localized in a central window of dijet invariant mass values that can in practice be varied to carry out a full search~\cite{Nachman:2020lpy, Hallin:2021wme, Buhmann:2023acn}. Instead, the unsupervised anomaly detection based on autoencoders employed in this work can be trained on the entire range of dijet invariant mass values as long as the signal fraction is sufficiently small. The autoencoder is trained to minimize the reconstruction loss resulting in a large anomaly score for BSM signal events. Ref.~\cite{Collins:2021nxn} found that the autoencoder performance remains stable for signal-to-background ratios up to $S/B \lesssim 1\%$, a result we were able to confirm in our numerical studies.

The event generation is performed using \textsc{Pythia8}~\cite{Pythia2}, and detector effects are simulated with \textsc{Delphes} 3.4.1~\cite{delpes3}. In both cases, default configurations are used, without pileup or multiple parton interactions (MPI). The final-state jets are reconstructed using the anti-$k_T$ clustering algorithm~\cite{Cacciari:2008gp}, as implemented in \textsc{FastJet}~\cite{fastjetManual}. We refer the reader to section \ref{app:datasets} on the details of the preprocessing of the events. A jet radius of $R = 1.0$ is used with a transverse momentum cut of $p_T > 1.2$~TeV. The signal events feature two high-energy jets containing the decay products of the boosted $X$ and $Y$ particles. These jets form a dijet system with large invariant mass, originating from the heavy $Z’$ boson. In addition, they exhibit a non-trivial two-pronged jet substructure. Background events are simulated using Standard Model processes that contain two jets satisfying the same transverse momentum cuts. For training our graph autoencoder, we use $10^5$ events with a tunable ratio of signal and background events. For validating the reconstruction loss, we use $10^4$ events with the same $S/B$ ratio as in training. Finally, to evaluate the model’s ability to distinguish between unseen background and anomalous events using a reconstruction loss threshold, we use $5 \cdot 10^4$ events that are not part of the training or validation sets. We verified that increasing the training sample size beyond the chosen value leads only to marginal performance improvements, while significantly increasing the computational cost, particularly when scanning over multiple graph architectures.

\subsection{Graph autoencoder}
\label{section:arch_anomaly}

We develop a GNN-based autoencoder for the anomaly detection task described above. The model takes a graph representation of individual jets as input, with different information assigned to both the graph nodes and edges. Each particle or subjet inside the main jet corresponds to a node. Instead of using the absolute positional information of the particles or subjets in the rapidity–azimuth plane, we only use the transverse momentum $p_{T,i}$ of each particle or subjet $i$ as a node input feature. The edges encode relationships between particles, in particular, their relative distances in the rapidity–azimuth plane. For each edge connecting particles $i$ and $j$, we include the following features $e_{ij}$~\cite{Qu:2022mxj}:
\begin{align}
\begin{split}
    \theta_{ij} &= (\Delta\eta_{ij}^2 + \Delta\phi_{ij}^2)^{1/2}\,, \\
    k_{T,ij} &= \min(p_{T,i}, p_{T,j}) \, \theta_{ij}\,, \\
    z_{ij} &= \min(p_{T,i}, p_{T,j}) / (p_{T,i} + p_{T,j})\,. \label{eq:interaction_gae}
\end{split}
\end{align}
Since jets have a variable number of constituents, each jet corresponds to a graph with a different number of nodes and edges. By contrast, exclusive $k_T$ subjet reclustering always produces a graph with a fixed number of nodes and edges. The graph autoencoder developed in this work is compatible with both fixed and variable length input.

A GNN-based autoencoder is a machine learning architecture designed to learn a low-dimensional representation of graph-structured data in an unsupervised way. It consists of two parts: an encoder, which processes the input graph through multiple message-passing layers (described in more detail below) to produce a compressed latent embedding; and a decoder, which reconstructs aspects of the original graph from this embedding. The model is trained by minimizing a reconstruction loss that encourages it to capture the essential features of the input data. A key feature of autoencoder-based anomaly detection is that its performance does not depend on the fraction of anomalous events in the signal region, as long as the model is trained on a background-rich sample. During training, the autoencoder learns to compress and reconstruct background events. When applied to test data, the reconstruction loss — or a related scoring function — serves as an anomaly score, flagging events that deviate significantly from the learned background distribution. A higher score indicates a larger deviation, which has been found to be largely independent of the overall signal fraction~\cite{Collins:2021nxn}.

\begin{figure*}[t]
\begin{center}
\centerline{\includegraphics[width=0.95\textwidth]{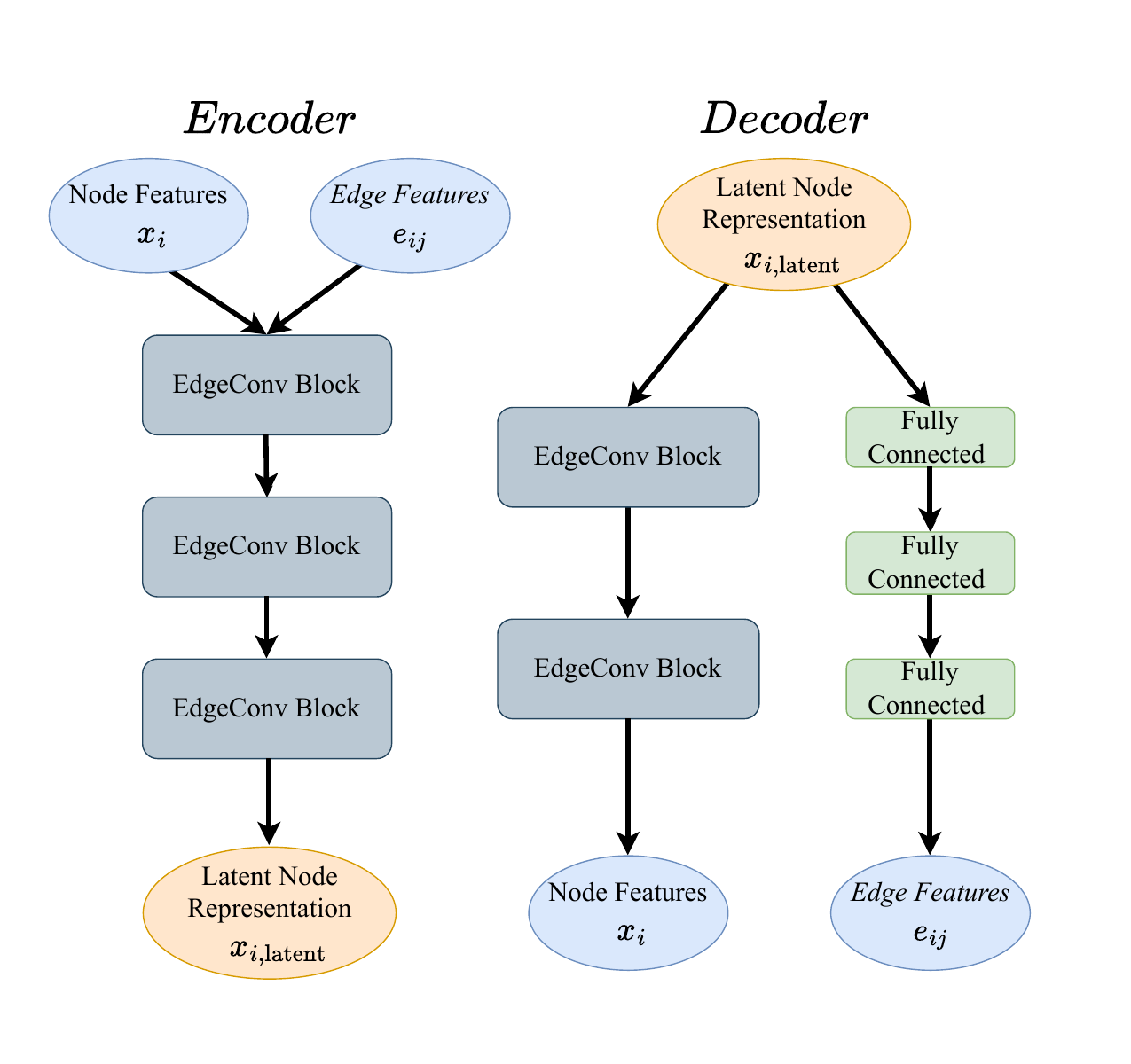}}
\vskip -0.2in
\caption{Schematic illustration of the GNN-based autoencoder developed in this work. For visualization purposes, we do not include the residual connections that exist within the encoder and the decoder. See text for more details.}
\label{fig:gae_arch}
\end{center}
\vskip -0.4in
\end{figure*}

Fig.~\ref{fig:gae_arch} shows a schematic illustration of the GNN-based autoencoder developed in this work. The encoder consists of a stack of custom message-passing blocks built upon the edge convolution (EdgeConv) layer introduced in Ref.~\cite{wang2019dynamic}. Each encoder block is designed to update the node features $x_i$ by incorporating information from the neighboring node features $x_j$ and the corresponding edge features $e_{ij}$. These blocks do not update the edge features. Instead, they reuse the same three relative features from the input layer as listed in Eq.~(\ref{eq:interaction_gae}). Each block is composed of two linear layers, both using a 64-dimensional auxiliary space for the node features. In the final block, the second layer outputs a lower-dimensional representation $x_{\rm latent}$ forming the bottleneck of the autoencoder. For each node $i$, the message-passing operation of layer $\ell$ is given by
\begin{equation}
    {x}_i^{\ell+1} = \text{ReLU}\sum_{j}\left( W_{e,1}^\ell \cdot e_{ij} + W_{e,2}^\ell \cdot x_j^\ell\right) \,,
\end{equation}
where we sum over the neighbors $j$ of node $i$. The matrices $W_{e,k}^\ell$ of the encoder with $k=1,2$ contain trainable weights, which are shared across all nodes and edges but vary from layer to layer. In addition, a residual connection~\cite{he2016deep} is applied after each block by adding the original node features $x_i^\ell$ to the updated representation. This helps to preserve important node-level information, ensures better gradient flow, and enables deeper network architectures. The sequence of encoder blocks maps the input graph to a lower-dimensional latent node representation ${x}_{\rm latent}$ with dimension $d_{\text{latent}}$. This latent representation focuses solely on node-specific information and does not retain explicit edge features.  A hyperparameter scan showed that $d_{\text{latent}} = 2$ provides the best performance, creating a tight bottleneck that ensures only a compressed summary of the original information is passed to the decoder. Although the latent space is defined per node, the input information is distributed between node features and edge features. For a graph with $N$ nodes and $2N-3$ edges, this corresponds to approximately $3N$ input degrees of freedom, even for the sparsest Laman graph structures considered here. Since the latent representation is restricted to node variables, a three-dimensional latent space would not constitute a bottleneck. A one-dimensional latent space was found to be insufficient to reconstruct the jet structure, while a two-dimensional latent space provides the minimal capacity required to capture the relevant features, corresponding to a compression factor of roughly $2/3$. The compression factor is more significant for unique-$k$ graphs. We therefore refer to this choice as a tight bottleneck.

The decoder architecture consists of two independent branches that operate on the latent node features $x_{\text{node}}$. The first branch reconstructs the node features and consists of a stack of decoder blocks that are structurally analogous to those in the encoder. The second branch reconstructs the edge features. A key difference in the decoder is the absence of edge inputs: the decoder blocks form messages by combining the features ${x}_i$ of node $i$ with the features of the neighboring nodes ${x}_j$ by concatenating them element-wise as ${x}_i - {x}_j$, as in the original EdgeConv proposal~\cite{wang2019dynamic}. The message-passing operation for the decoder block is given by
\begin{equation}
    x_i^{\ell+1} =  \text{ReLU}\sum_{j}\left( W_{d,1}^\ell \cdot (x_i^\ell - x_j^\ell) + W_{d,2}^\ell \cdot x_i^\ell \right)\,,
\end{equation}
with trainable weights matrices $W^\ell_{d,k}$ analogous to the encoder operation. Each decoder block uses two linear layers of dimension 64. This branch ultimately maps the latent node features back to the original 1D node inputs. As in the encoder, residual connections are added to help preserve information and support deeper architectures. The second decoder branch is responsible for reconstructing the edge features. It takes the latent representations of each pair of connected nodes (${ x}_i, {x}_j$) and processes them through a symmetric function. Specifically, we use the concatenation of element-wise minima and maxima. The output is passed through two fully connected layers (each of dimension 64) followed by a final linear layer that produces a 3D output, matching the dimensionality of the input edge features. A residual connection is also added in this branch to ensure stable propagation of $x_{\rm latent}$ through the network.

The model is trained end-to-end by minimizing a combined loss function, for both node and edge features. We use Mean Squared Error (MSE) as the base loss function, following prior studies showing its suitability for anomaly detection~\cite{tsan2021particlegraphautoencodersdifferentiable}. The total loss that the model is trained on is the sum of the MSE loss for the node and edge features
\begin{equation}
    \mathcal{L} = \text{MSE}_{\rm nodes} +  \text{MSE}_{\rm edges}\,.
\label{eq:recon_loss}
\end{equation}
Here, the MSE loss is defined by summing over all vertices and edges, respectively,  
\begin{equation}
\text{MSE}_{\rm nodes} = \sum_{v \in V} | x_v - \hat{x}_v |^2\,,\quad\text{MSE}_{\rm edges} = \sum_{{ij} \in E} | e_{ij} - \hat{e}_{ij} |^2 \,.
\end{equation}
The hatted quantities denote the original inputs the autoencoder is trained to reconstruct. Since the input node features consist only of the transverse momenta $p_{Ti}$ of the particles or subjets, while the latent node dimension is two-dimensional, the autoencoder can relatively easily reconstruct the transverse momenta. As a result, the reconstruction loss is dominated by the edge features. Potentially, a relative scaling factor between the two contributions to the total loss in Eq.~(\ref{eq:recon_loss}) could be used to further improve the performance but we do not explore this option here. Since the LHC Olympics anomaly detection data set contains dijets, we independently process the two jets through the autoencoder. Their respective reconstruction loss is summed to obtain the total event-level loss. An alternative approach, which is left for future work, would be to train separate models for the leading and subleading jets, or to apply thresholds to each jet’s reconstruction loss individually rather than summing them.

After training, we quantify the performance of the autoencoder by evaluating its ability to distinguish between background and signal in terms of differences in the reconstruction loss. Specifically, we consider the Area Under the Curve (AUC) of the Receiver Operator Characteristic (ROC) curve. The ROC curve is the cumulative distribution function of the true positive rate as a function of the false positive rate of a binary classification task as the decision threshold is varied. In addition, we consider a more relevant metric for anomaly detection: the maximum value of the Significance Improvement Characteristic (SIC) curve~\cite{Gallicchio:2010dq}. The SIC quantifies the gain in statistical significance achieved by applying a cut on the classifier output score. For a given threshold, the signal efficiency (true positive rate), $\epsilon_S$, is the fraction of signal events that pass the selection, while the background efficiency (false positive rate), $\epsilon_B$, is the fraction of background events that also pass the cut. The significance improvement for the cut is then given by $\text{SIC} = \epsilon_S/\sqrt{\epsilon_B}$. In our case, the anomaly score is the reconstruction loss; see Fig.~\ref{fig:recon_loss_anomaly}. By varying the decision threshold, we determine which events are classified as background or signal. To estimate the statistical uncertainty in the performance metrics, we train the neural networks four times for each task and use the standard deviation of the resulting scores as a proxy for model uncertainty.

We note that we do \textit{not} use any batch normalization layers in either the encoder or the decoder as we find that they degrade the anomaly detection performance by up to $\approx 20 \%$ in terms of the background rejection rate, or similarly, $1-\text{AUC}$. Batch normalization is typically used to mitigate internal covariate shift~\cite{ioffe2015batchnormalizationacceleratingdeep} by computing the batch-wise mean $\mu$ and variance $\sigma^2$, and normalizing the data via $x \rightarrow (x-\mu)/\sigma$. However, this normalization can inadvertently smooth out the tail regions of the data distribution. In anomaly detection, where the test set may include both in-distribution (background) and out-of-distribution (signal) samples, this smoothing reduces the contrast between signal and background, weakening the model’s ability to detect anomalies. While the overall reconstruction loss may be reduced, the discriminatory power is diminished.

We implement the model in \texttt{PyTorch}~\cite{paszke2019pytorch} using the AdamW optimizer~\cite{kingma2014adam}. We train the model for 50 epochs using a batch size of 512 and a linearly annealed learning rate. The learning rate starts at a peak value of $3 \cdot 10^{-3}$ and decreases to $2 \cdot 10^{-4}$ by the end of the training.  During training, the model minimizes the reconstruction loss on the background-rich data set. During testing, it distinguishes signal from background by applying a threshold to the total reconstruction loss, as shown in Fig.~\ref{fig:recon_loss_anomaly}.

\subsection{Numerical results}
\label{section:anomaly_detection}

\begin{figure*}[t]
\centering
\includegraphics[width=0.7 \textwidth]{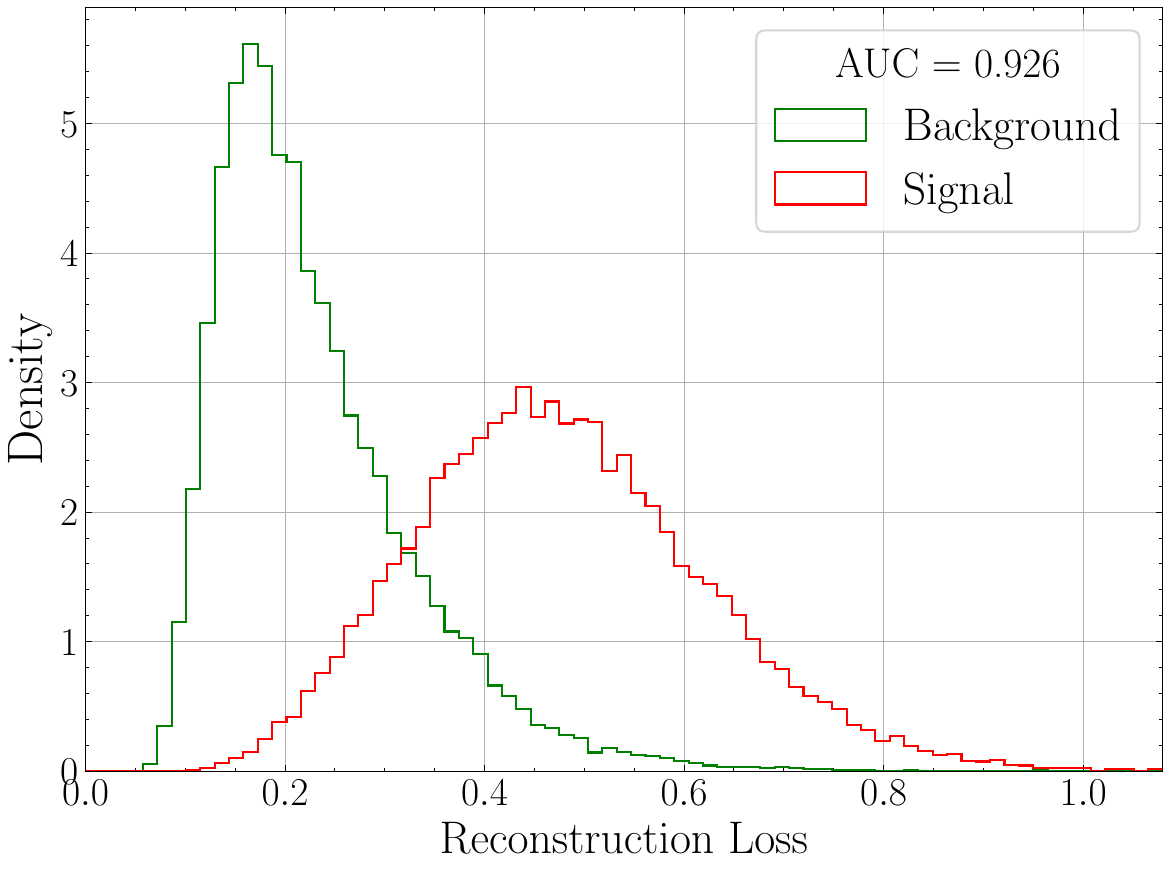}
\caption{The reconstruction loss of the graph autoencoder for a representative run based on a sparse \textit{unique-6} graph and $n_{\rm subjets}=30$.} 
\label{fig:recon_loss_anomaly}
\end{figure*}

As described in section~\ref{sec:anomaly_dataset}, the LHC Olympics data set contains dijet background events simulated using Standard Model processes, along with signal events corresponding to the resonant production of a new heavy boson~\cite{lhco}. The new particle has a mass of 3.5~TeV, producing a localized bump in the dijet invariant mass spectrum $m_{jj}$. When the signal fraction is small relative to the background, the new particle is difficult to identify using traditional bump-hunting techniques. However, machine learning-based anomaly detection methods that exploit full event-level information can successfully identify such signals. As described in the previous section, the anomaly detection strategy employed in this work relies on an autoencoder trained in an unsupervised manner, using sparse graph representations of jets. The training data set for the autoencoder may include both signal and background events, provided the signal fraction remains sufficiently small. We now present numerical results obtained by applying the autoencoder described above to the LHC Olympics data set, on the full range of the dijet mass $m_{jj}$, using a variable number of reconstructed $k_T$ subjets and exploring a range of graph construction strategies. For the results shown here, we use a signal-to-background ratio of $S/B = 3\%$. However, we note that comparable performance is observed across a range of values with $S/B \leq 3\%$, consistent with expectations~\cite{Collins:2021nxn}. An illustrative result for the reconstruction loss at test time is shown in Fig.~\ref{fig:recon_loss_anomaly}. In this example, we use $n_{\rm subjets} = 30$ exclusive subjets and employ the \textit{unique-6} sparse graph construction, as described in section~\ref{sec:graphconstruction} above. By scanning over different decision thresholds, we obtain the corresponding ROC curve which determines the AUC and SIC, which are our chosen performance metrics for evaluating the various algorithms considered in this section. 

\begin{figure*}[t]
    \centering 
    \begin{tabular}{c @{} c} 
        \begin{minipage}{0.49\textwidth} 
            \centering
            \includegraphics[width=\linewidth]{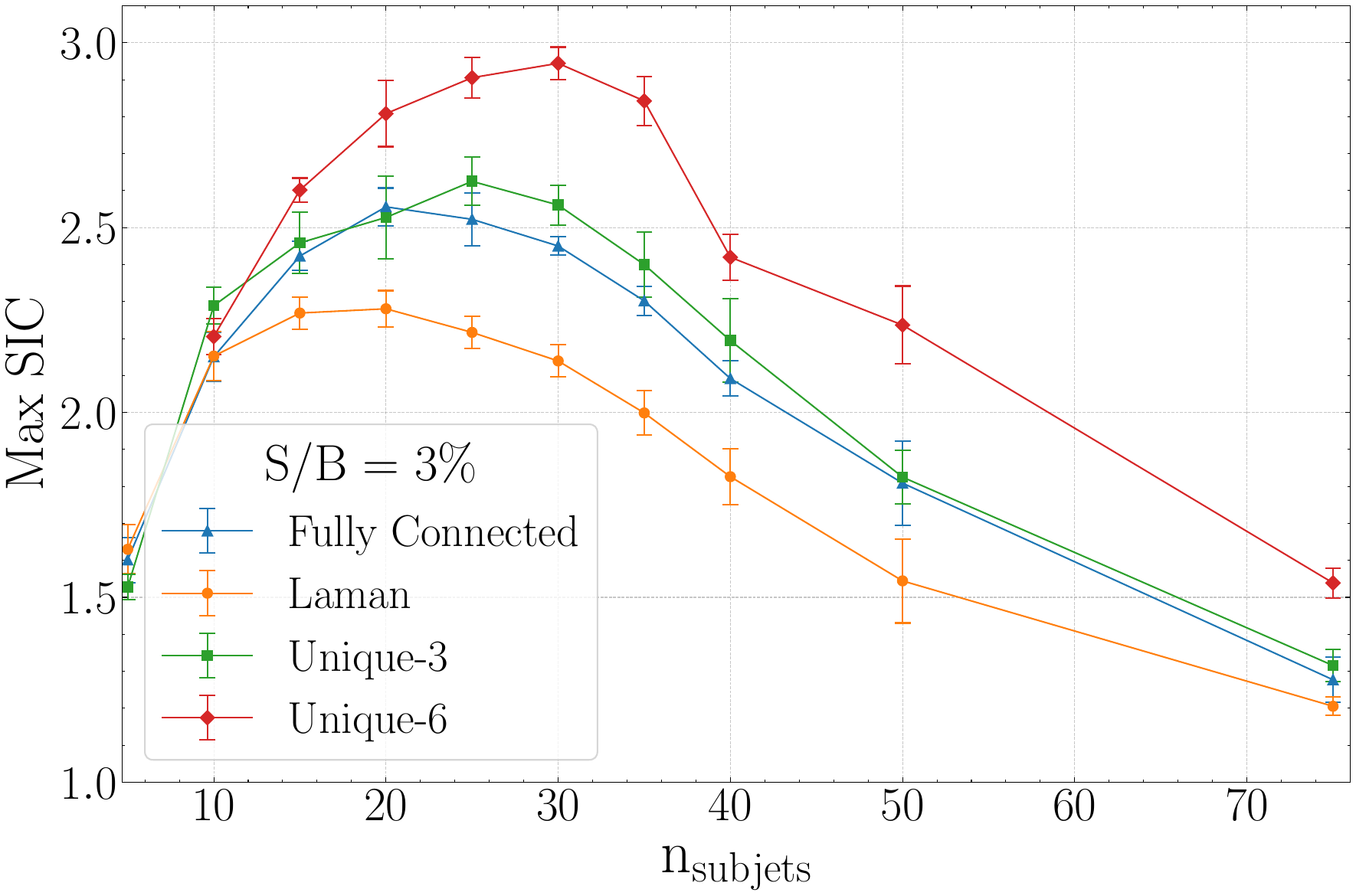}
            \label{fig:auc_anomaly}
        \end{minipage}
        \begin{minipage}{0.49\textwidth} 
            \centering
            \includegraphics[width=\linewidth]{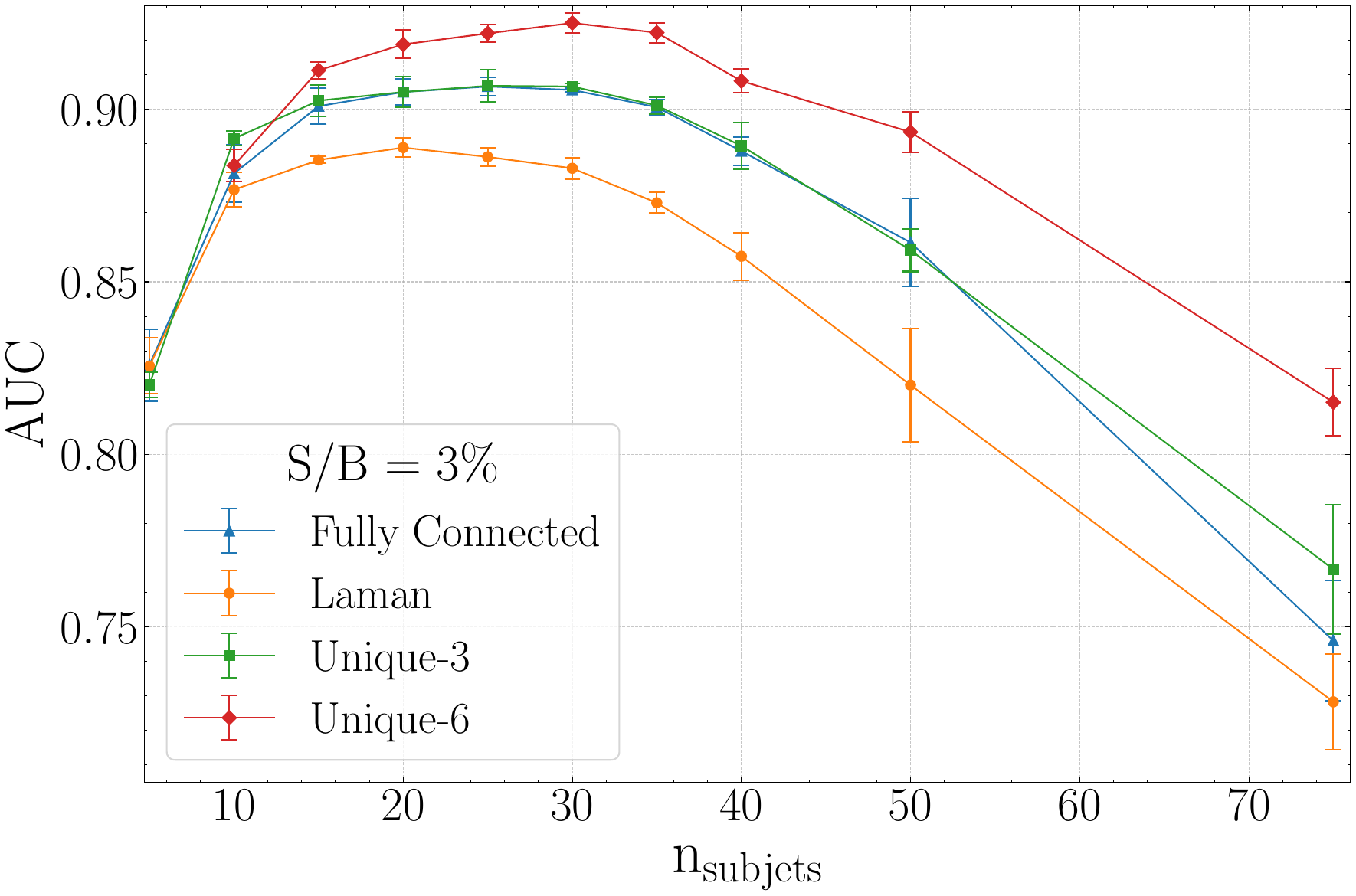} 
            \label{fig:auc_anomaly_nsub30}
        \end{minipage}
        \\ 
    \end{tabular}
    \caption{Anomaly detection performance (left: Maximum SIC, right: AUC) of our graph autoencoder as a function of the number of exclusive subjets for different graph constructions. The two curves exhibit similar behavior, with both performance metrics peaking for the \textit{unique-6} graph at $n_{\rm subjets} = 30$, resulting in a max SIC of 2.94 and an AUC of 0.925, respectively.}
    \label{fig:relgae_results}
\end{figure*}

\begin{figure*}[t]
\centering
\includegraphics[width=0.7 \textwidth]{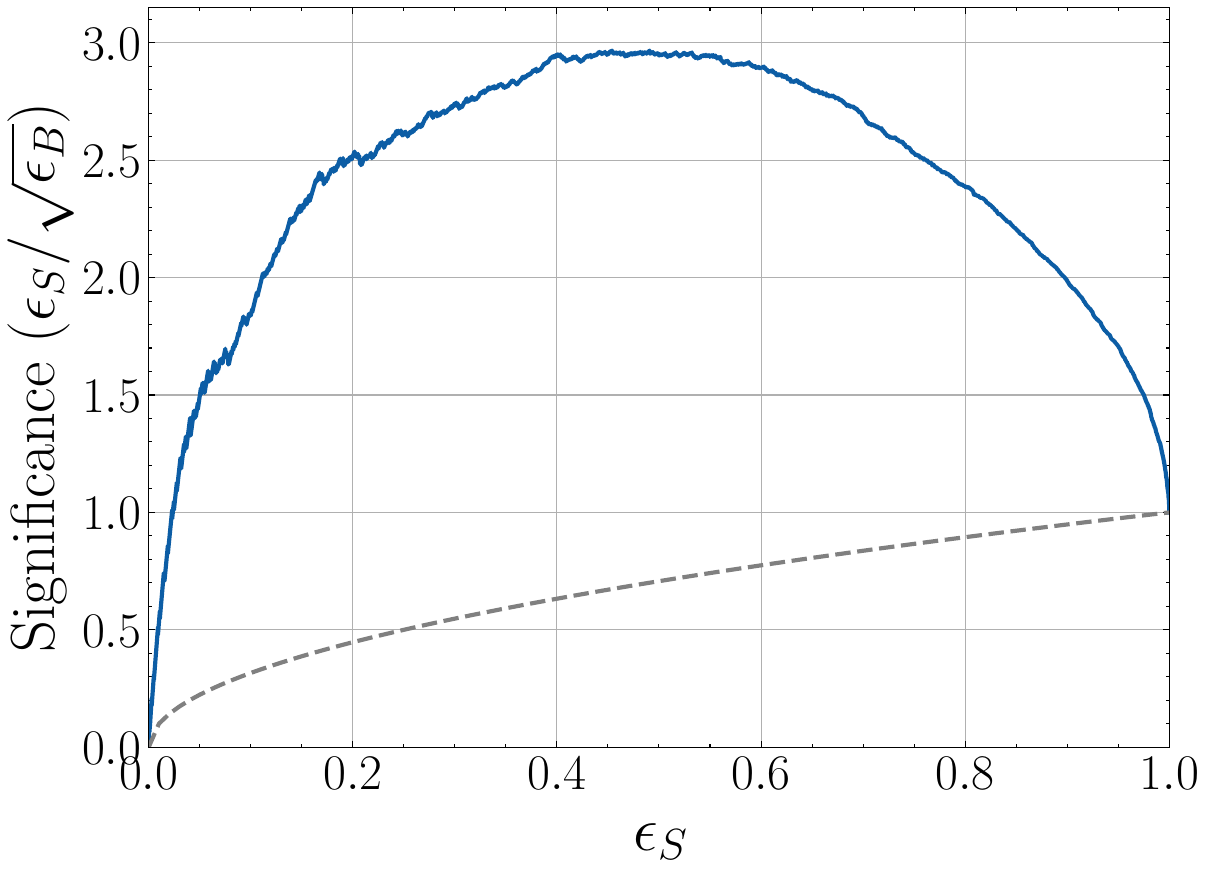}
\caption{The SIC curve based on a sparse \textit{unique-6} graph for $n_{\rm subjets}=30$. The dashed line indicates the behavior of a random classifier with $\epsilon_S = \epsilon_B$. The SIC corresponds to the same run as shown in Fig.~\ref{fig:recon_loss_anomaly}.} 
\label{fig:sic}
\end{figure*}

The results for the performance of our graph-based autoencoder are shown in Fig.~\ref{fig:relgae_results}, which represent the central findings of our study. Each numerical value corresponds to the average over four independent runs, and the error bars indicate the standard deviation across these runs. To ensure the robustness of our results, we monitored the training convergence. In a small fraction of instances (approximately 5\% on average), the training process did not converge to a satisfactory minimum for the reconstruction loss. These runs, identified by a significantly higher training loss, i.e. not within 75\% of the best loss achieved by the other runs of that data point, were discarded and repeated until four successfully converged runs were obtained for each reported data point. In Fig.~\ref{fig:relgae_results}, we show the max SIC (left panel) and the AUC (right panel) for fully connected (blue), Laman (orange), {\it unique-6} (red), and {\it unique-3} (green) graphs. For completeness, we show the SIC for a representative run in Fig.~\ref{fig:sic}.

First, we find that the performance of the graph-based autoencoder peaks at intermediate values of $n_{\rm subjets}$ across all tested graph constructions. This behavior contrasts with typical jet classification tasks, where the performance peaks at large values of $n_{\rm subjets}$ or directly at the hadron level, i.e., without any subjet clustering. For further comparison, see Appendix~\ref{section:appendix}, where we reproduce this trend using various graph-based jet classification methods. This is consistent with the findings of Ref.~\cite{Athanasakos:2023fhq}, which used a deep sets-based classifier based on subjets. These findings highlight a difference between supervised classification and unsupervised anomaly detection: while classification generally benefits from richer representations, the latter may suffer from excessive input complexity, which can lead to overfitting. Subjet clustering, therefore, acts as a useful physics-informed bottleneck that can guide the model toward physically meaningful structures, resulting in improved performance. Rather than hand-selecting a few high-level observables, we can smoothly interpolate between high- and low-level information by adjusting the number of subjets $n_{\rm subjet}$, achieving optimal performance in the range $n_{\rm subjet} \approx 25-30$. 

Second, we observe that autoencoders based on fully connected graphs, which encode the maximal amount of information, do not yield the best performance. While Laman graphs tend to underperform for $n_{\rm subjet} \gtrsim 10$, the \textit{unique-3} graph performs comparably or better, within the shown error bars, than the fully connected graph across all values of $n_{\rm subjet}$. The sparse \textit{unique-6} graphs consistently achieve the best performance among all \textit{unique-k} graphs, which peaks for $n_{\rm subjet} \sim 25-30$, in Fig.~\ref{fig:relgae_results}. A similar behavior, when comparing against the fully connected graph, was observed for higher-point unique graph constructions as well. The relatively low performance of Laman graphs may be due to the fact that edge information does not uniquely determine the graph structure, in contrast to the case of unique graphs. This suggests that, in addition to subjet reconstruction, sparse but structured graphs can serve as effective inductive biases that retain sufficient geometric structure allowing for an efficient learning procedure. 

\begin{figure*}[t]
\centering
\includegraphics[width=0.8 \textwidth]{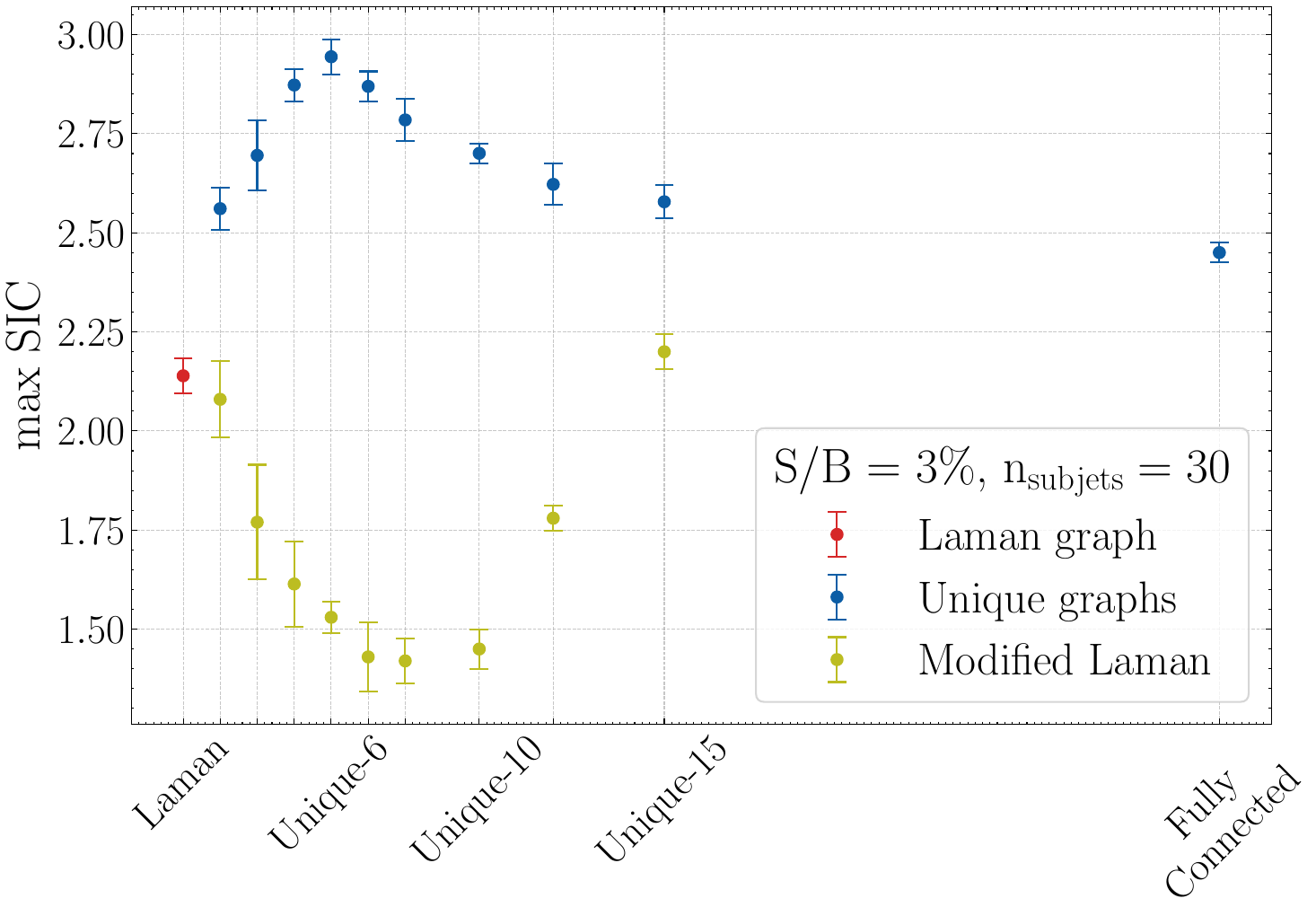}
\caption{Maximum value of the SIC curve as a function of the connectivity for different graph constructions, for a fixed number of subjets $n_{\rm subjets} = 30$. Here the \textit{unique-30} graph is fully connected. The highest max SIC value is achieved for {\it unique-k} graphs with an intermediate level of sparsity.}
\label{fig:maxsic_nsub30}
\end{figure*}

Next, in Fig.~\ref{fig:maxsic_nsub30}, we plot the maximum SIC value for different graph constructions for $n_{subjets}=30$ that corresponds to the best performance for most of the unique graphs, see Fig.~\ref{fig:relgae_results}. We show the result for the Laman graph (red), several unique graphs and the fully connected graph (blue). We observe a peak at intermediate levels of sparsity with $\text{max SIC} \sim 3$ around the \textit{unique-6} graph. Adding significantly more edges ultimately leads to a drop in performance. Similar to Fig.~\ref{fig:relgae_results}, this result indicates that systematically increasing the sparsity of the graph-based autoencoder can lead to improved performance. 

To further investigate the impact of graph connectivity and to disentangle the effects of sparsity from those of graph topology, we introduce an alternative graph construction algorithm. The yellow markers in Fig.~\ref{fig:maxsic_nsub30} show the performance of this alternative algorithm, which interpolates between Laman and fully connected graphs analogous to the \textit{unique-k} graphs. This construction is referred to as ``modified Laman'' in Fig.\ref{fig:maxsic_nsub30}. Starting from a Laman graph, we successively add edges by computing the angular distances in the $\eta$-$\phi$ plane between all pairs of subjets that are not yet connected. Edges are then added one-by-one between the closest unconnected pairs until the graph becomes fully connected. The yellow markers in Fig.~\ref{fig:maxsic_nsub30} show the performance of these modified Laman graphs as a function of the number of edges, which is directly compared to the corresponding \textit{unique-k} graphs shown by the blue markers. As both constructions eventually converge to the fully connected graph, their performance also becomes comparable for a large number of edges. However, while the unique graphs display a peak in performance at intermediate levels of sparsity, as discussed above, the performance of the modified Laman graphs decreases significantly for the same number of edges. This behavior suggests that it is not merely the degree of sparsity that explains the improved performance of \textit{unique-k} graphs, but rather their specific connectivity structure. We conclude that the graph-theoretic design of \textit{unique-k} graphs more effectively captures the topology of the underlying phase space. In fact, to the best of our knowledge, the performance of the \textit{unique-6} graph represents the current state-of-the-art among unsupervised autoencoder-based methods for this anomaly detection benchmark data set. In addition, we compared our results to sparse $k$-nearest-neighbor graphs. Although these constructions also yield sparse graphs, they generally do not satisfy the rigidity conditions, such as Laman or globally rigid graphs. As expected, we find that their performance is generally worse compared to that of fully connected graphs.

We also tested a graph autoencoder based solely on absolute positional (node) information~\cite{tsan2021particlegraphautoencodersdifferentiable}, without edge features, but it did not achieve competitive performance. Similarly, retaining the edge prediction while adding absolute node information and reconstructing the three momenta $(p_{Ti}, \eta_i, \phi_i)$, similarly to Ref.~\cite{Atkinson_2021}, led to a maximum value of the SIC of $\sim 2$. These findings are consistent with the expectation that the relevant physics information is primarily encoded in the relative relationships between particles or subjets. Potential improvements of our results could be obtained by modifying the latent space representation that acts as a bottleneck of the graph autoencoder. Currently, the latent representation only focuses on the node information, outputting a 2D vector for every particle or subjet, but any graph-related structural information is dropped. It has been noted in previous studies ~\cite{Batson:2021agz, Ngairangbam:2025fst} that the topological properties of both the data and the latent space can obstruct the reconstruction properties and the generalizability of an autoencoder. One way to improve this effect would be to project the starting graph, down to a sparse graph in the latent space, since this could match the manifold of the data. An exploration of this will be left for future work. 

Reconstruction-based autoencoders are known to exhibit a complexity bias, where topologically simple objects such as QCD jets are reconstructed with high fidelity even if they are underrepresented or entirely absent from the training data \cite{Finke:2021sdf}. This behavior is also observed in our setup and appears to be largely independent of the specific graph construction employed. Several recent works have proposed architectural modifications to mitigate this effect, including the use of normalized autoencoders~\cite{Dillon:2022mkq}. Exploring such approaches in combination with sparse and rigidity-motivated graph structures represents an interesting direction for future studies.

While the maximum SIC values achieved by autoencoder-based methods are often modest, this behavior is intrinsic to the paradigm and has been discussed in the literature. We note that here, we focus on relative performance improvements related to the graph structure, rather than on absolute sensitivity or discovery potential. A more detailed quantitative comparison with weakly supervised methods, particularly in the transition region of signal fractions, is left for future work.

\section{Conclusions and outlook}
\label{section:conclusions}

Anomaly detection plays an increasingly important role in searches for new physics at collider experiments, where the goal is to identify rare events without relying on detailed signal models. In this work, we explored how graph-theoretic insights can inform the design of machine learning architectures for this task, with a particular focus on graph autoencoders constructed from sparse graphs. Motivated by the use of relative information instead of absolute particle positions, we investigated different sparse graph construction techniques and their impact on performance for anomaly detection and jet classification tasks, in comparison to fully connected graphs.  

We found that the choice of graph used to represent a jet of particles can enhance the performance of machine learning models by embedding physically motivated inductive biases into the architecture. In particular, we focused on (1) locally rigid Laman graphs, and (2) different unique graph constructions that are globally rigid and admit only a single embedding in the rapidity–azimuth plane. As an exemplary application, we trained a graph-based autoencoder to reconstruct both particle features and pairwise geometric information using the data set from the LHC Olympics anomaly detection challenge. We observed that certain {\it unique-$k$} graphs outperform fully connected graphs, resulting in a higher maximum Significance Improvement Characteristic (SIC) curve. Our results show that not only the sparsity of the graph, but also its specific connectivity structure, is critical for improved performance. Additionally, we explored the effect of clustering particles into subjets, allowing for a smooth interpolation between low- and high-level representations. Similar to the use of relatively sparse graphs, we found that the autoencoder performance peaks at intermediate levels of clustering with $n_{\rm subjets}\approx 30$.

These results motivate further exploration of how graph-theoretic insights can enhance machine learning approaches in collider physics. In particular, the connection between graph theory and computational complexity may enhance the interpretability of machine learning algorithms and inform the development of architectures with improved performance. Further investigations are also needed to study the dependence of the anomaly detection performance on the type of signal. In addition, it would be interesting to investigate whether similar graph-based methods can benefit other anomaly detection strategies.

\section*{Code Availability}

The code developed for this work is available on GitHub.

\begin{itemize}
    \item The Graph Autoencoder for anomaly detection can be found at: \url{https://github.com/DimAthanasakos/Graph-AutoEncoder-for-Model-Agnostic-Anomaly-Detection}
    \item The code for jet classification, as described in the Appendix, is available at: \url{https://github.com/DimAthanasakos/Graphs-and-Jet-Classification}
\end{itemize}

\section*{Acknowledgments}

We would like to thank Andrew Larkoski and James Mulligan for collaborating during the early stages of this work. In addition, we would like to thank Joseph Mitchell, Vasilis Belis, and Nobuo Sato for helpful discussions. DA is supported by the NSF Grant PHY2210533 and in part by the Onassis Foundation. This research used resources of the National Energy Research Scientific Computing Center (NERSC), a DOE Office of Science User Facility using NERSC awards NP-ERCAP0033891 and NP-ERCAP0031584. MP is supported by the U.S. Department of Energy, Oﬃce of Science, Oﬃce of Nuclear Physics, under the contract DE-AC02-05CH11231. FR is supported by the DOE, Office of Science, Office of Nuclear Physics, Early Career Program under contract No. DE-SC0024358/DE-SC0025881. We thank the Institute for Nuclear Theory at the University of Washington for its kind hospitality and stimulating research environment. This research was supported in part by the INT's U.S. Department of Energy grant No. DE-FG02-00ER41132.

\appendix
\section*{Appendix}
\section{Applications to jet classification}

\label{section:appendix}

In this section, we apply the sparse graph constructions introduced in section~\ref{sec:graphconstruction} to different jet classification tasks. Unlike in the anomaly detection task discussed above, we do not observe an improvement in classification performance when using sparse graphs compared to fully connected graphs. Nevertheless, exploring different graph construction techniques can provide useful insights. Here, we focus only on graph constructions based on the particles inside the jet, since jet classification using subjets has already been discussed in Ref.~\cite{Athanasakos:2023fhq} in the context of deep sets.

\subsection{Data sets}\label{app:datasets}

We consider three representative binary jet classification tasks in high-energy physics: quark vs. gluon jets, $Z$ vs. QCD jets, and top vs. QCD jets. See Refs. for more details~\cite{Lonnblad:1990bi,deOliveira:2015xxd, Komiske:2016rsd, Komiske:2017aww, Kasieczka:2017nvn, Louppe:2017ipp, Komiske:2018cqr, Qu:2019gqs, Kasieczka:2020nyd, Dreyer:2020brq, Top_taggers_review, Butter:2017cot, Chen:2019uar,Araz:2021wqm, Gong:2022lye, Schwartz:2021ftp,Khot:2022aky, Lin:2018cin, Khosa:2021cyk, Nakai:2020kuu, Bogatskiy:2023nnw, CMS:2020poo, EnergyFlowPolynomials:2017aww, Romero:2021qlf, Konar:2021zdg, Larkoski:2019nwj, Choi:2018dag, Bogatskiy:2020tje, Esmail:2025kii,Krause:2025qnl}. For the quark vs. gluon classification task, we use the data set provided in Ref.~\cite{Zenodo:EnergyFlow:Pythia8QGs}, which consists of 2M jets with transverse momenta $p_T \in [500, 550]$~GeV, rapidity $|\eta| < 1.7$. Jets are clustered using the anti-$k_T$ algorithm~\cite{Cacciari:2008gp} with radius parameter $R = 0.4$. The jets were generated with \textsc{Pythia8}~\cite{Sjostrand:2014zea} at a center-of-mass energy of $\sqrt{s} = 14$~TeV. For the $Z$ vs. QCD and top vs. QCD classification tasks, we use the JetClass data set of Refs.~\cite{Qu:2022mxj, qu2022jetclass}. The signal processes for these tasks correspond to boosted $Z \rightarrow q \bar{q}$ and $t \rightarrow b q q’$ decays, respectively. The background QCD jets are simulated using processes such as $q \bar{q} \rightarrow Z(\rightarrow \nu \bar{\nu}) + g$ and $q \bar{q} \rightarrow Z(\rightarrow \nu \bar{\nu}) + (uds)$. The simulation was carried out with \textsc{Pythia8}~\cite{Pythia2}, while detector effects are modelled using \textsc{DELPHES}~\cite{de2014delphes} with the CMS detector card. The jet reconstruction is performed using the anti-$k_T$ algorithm with $R = 0.8$, transverse momentum $p_T \in [500, 1000]$~GeV, and rapidity $|\eta| < 2$. For all three tasks, we use 2M jets. Each data set is split into training, validation, and test sets using an 80$\%, 10\%, 10\%$ split.

\subsection{Graph neural network classifier}

To embed the different graph structures into a classifier and compare with current state-of-the-art results, we use a modified version of the Particle Transformer (\texttt{ParT}) classifier~\cite{Qu:2022mxj}. In the original \texttt{ParT}, there are two inputs: the \textit{particle} input, which includes features for each particle, and the \textit{interaction} input, which captures features of particle pairs. As in the anomaly detection task, we aim to restrict the machine’s access to relative distances rather than absolute positions in the rapidity–azimuth plane. Therefore, we limit the {\it particle} input of the \texttt{ParT} architecture to the transverse momenta of all particles inside the jet. In addition, we include non-zero {\it interaction} features in \texttt{ParT} only for particle pairs connected by an edge in the corresponding graph construction~\cite{ying2021transformers}. In contrast to the GNN used in the graph autoencoder discussed in section~\ref{section:arch_anomaly}, every particle pair in the \texttt{ParT} architecture is connected via an attention layer that uses node features, but does not access their relative distances unless it is included in the input graph. The pairwise features include logarithmic versions of all the variables listed in Eq.~(\ref{eq:interaction_gae}), as well as
\begin{align}
    m^2 = (E_i+E_j)^2 - |\vec p_{i}+\vec p_{j}|^2\,.
\label{eq:interaction}
\end{align}
In the limit of highly boosted jets, $m^2$ depends only on the pairwise angles $\theta_{ij}$ and the transverse momenta $p_{Ti}$. These modifications allow us to use the \texttt{ParT} architecture while incorporating the modified graph structure.

The initial particle embedding is processed sequentially through a stack of particle attention blocks, which refine the embeddings via a multi-head self-attention mechanism. The interaction matrix acts as a bias term for the pre-softmax attention weights and is applied identically across all particle attention blocks, effectively serving as a residual connection~\cite{he2016deep}. The final embedding is then passed to two class attention blocks. The resulting class token is processed through a single-layer MLP and a softmax layer to generate the final classification scores. We implement the architectures in \texttt{PyTorch}~\cite{paszke2019pytorch}, using 8 particle attention blocks and 2 class attention blocks. The training is performed for 20 epochs, with a batch size of 512, using the AdamW optimizer~\cite{kingma2014adam} and a linear annealing scheduler with a warm-up phase. The initial learning rate is $2 \cdot 10^{-4}$, which linearly increases over the first 2 epochs to a peak of $2 \cdot 10^{-3}$, and then it linearly decreases to $4 \cdot 10^{-5}$.

\subsection{Numerical results and interpretability}

\begin{table}[t]
\centering
\begin{tabular}{|c|c|c|c|c|}
\hline
\textit{\textbf{AUC}} & \textit{Laman } & \textit{unique-3}& \textit{ParT ($p_T, \eta, \phi,\theta_{ij}$)} & \textit{ParT ($p_T,\theta_{ij}$)}  \\ \hline
\textit{q vs. g}& 0.8989 $\pm$ 0.0007 & 0.9007 $\pm$ 0.0005 & {0.9032} $\pm$ {0.0009} & {0.9041} $\pm$ \textit{0.0006} \\ \hline
\textit{Z vs. QCD}& 0.9439  $\pm$ 0.0006 & 0.9460 $\pm$ 0.0003 & {0.9483} $\pm$ {0.0006} & {0.9480} $\pm$ \textit{0.0007} \\ \hline
\textit{t vs. QCD}& 0.9829 $\pm$ 0.0004 & 0.9841 $\pm$ 0.0003 & {0.9865} $\pm$ {0.0005} & {0.9862} $\pm$  \textit{0.0005} \\ \hline
\end{tabular}
\caption{The performance of the different classifiers quantified in terms of the AUC for the three discrimination tasks: quark vs. gluon, $Z$ vs. QCD, and top vs. QCD jets.}
\label{table:auctable}
\end{table}

\begin{figure*}[h]
\centering
\includegraphics[width=0.95 \textwidth]{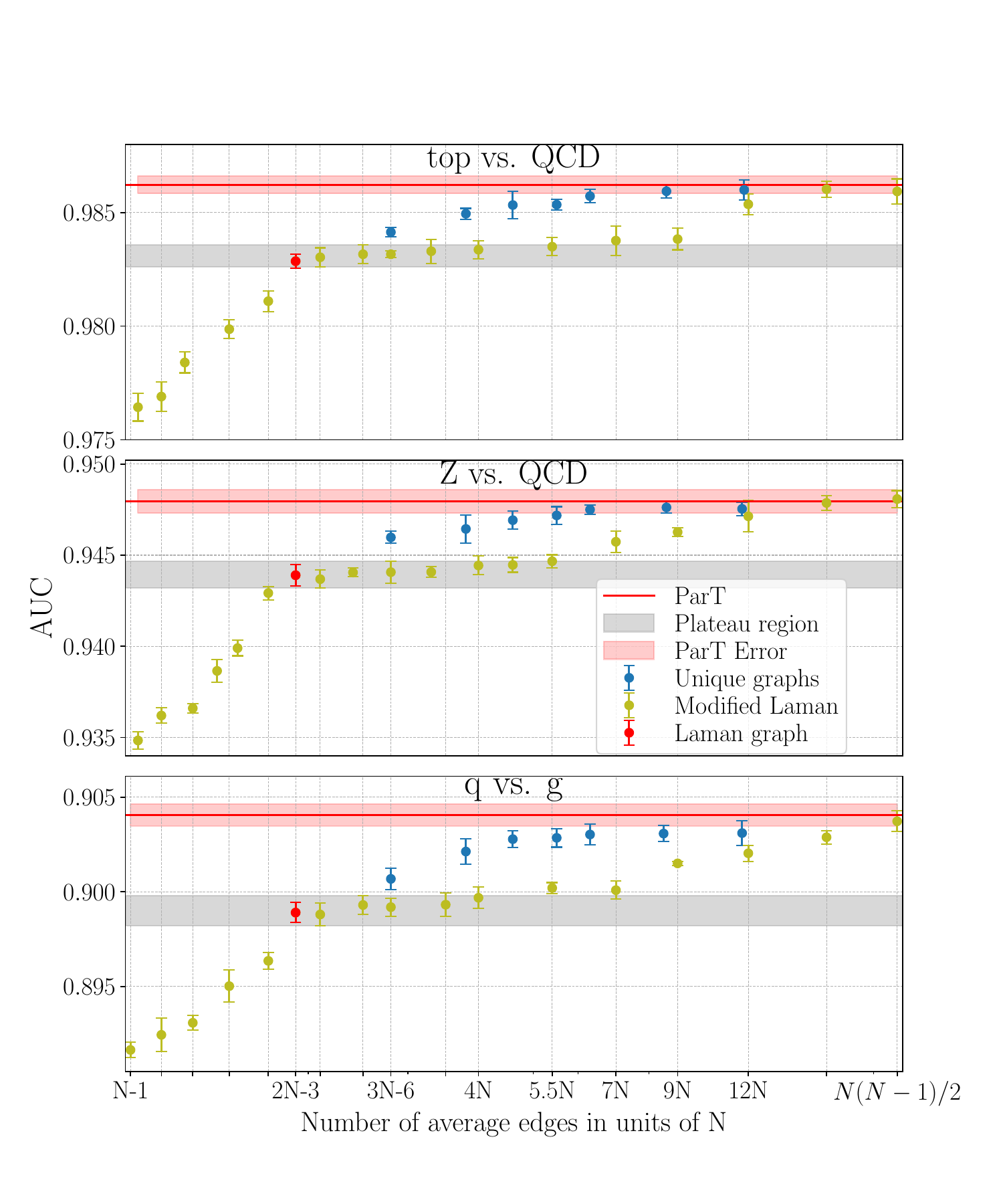}
\caption{Performance of different graph-based jet classifiers. The obtained AUC is shown as a function of the average number of edges that the input graph has in units of the particle number $N$ inside a jet (log scale). From top to bottom, the results are shown for: top vs QCD, $Z$ vs QCD, and quark vs. gluon jets.}
\label{fig:auc_classification}
\end{figure*}

We assess the performance of the classifier for the three binary classification tasks using the obtained AUC as the metric. Statistical uncertainties are estimated by training the model four times for each task and computing the standard deviation across these runs. The AUC results for all three classification tasks are summarized in Table~\ref{table:auctable}. In the first two columns, we report the results of the modified \texttt{ParT} classifier using Laman and \textit{unique-3} graph constructions, respectively. The third column shows the performance of the original \texttt{ParT} classifier, where each node includes full particle information $(p_T, \eta, \phi)$. In the fourth column, the node information is restricted to only the transverse momentum $p_T$, while the classifier is provided with relative angular distances $\theta_{ij}$ between all particle pairs. First, we observe that the performance of \texttt{ParT} using either absolute positions or only relative distances agrees within the estimated uncertainties for all three classification tasks. This observation is in line with the expectation that the relevant physics of jet classification is encoded in the relative information of particles instead of absolute positions. Next, while the performance gap between the Laman graph-based classifier and the original \texttt{ParT} is statistically significant, the AUC values differ by only about $0.3-0.4\%$. This is a remarkably small difference, given that the Laman graph-based classifier has access to roughly an order of magnitude fewer pairwise distances $\theta_{ij}$. For an average jet particle multiplicity of $\langle N \rangle \approx 50$ at the LHC, the average ratio of pairwise angles in a fully connected graph to those in a Laman graph is approximately $(N(N-1)/2)/(2N-3) \approx 12$. The classifier based on the \textit{unique-3} graph outperforms the one using Laman graphs, though a small but statistically significant difference remains compared to the original \texttt{ParT}. However, as we increase $k$ in the \textit{unique-$k$} graph constructions, this performance gap closes, as illustrated in Fig.~\ref{fig:auc_classification}.

To illustrate that the specific choice of graph construction impacts how quickly the performance gap closes as the number of edges increases, we show in Fig.~\ref{fig:auc_classification} the AUC for all three classification tasks as a function of the average number of edges in the input graph. Similar to Fig.~\ref{fig:maxsic_nsub30}, we compare the \textit{unique-k} graphs to the modified Laman graphs. Here, the starting point is the Laman graph, indicated by the red marker. The yellow markers represent the AUC performance when edges are either added to or removed from the Laman graph. To add edges, we calculate the pairwise distances of all particles in a jet that are not yet connected, and we progressively connect the nearest particles with an edge. This process continues until the graph becomes fully connected. To remove edges, we randomly select a node and delete the edge connecting it to the hardest particle. We ensure that the entire graph is still connected. Therefore, if removing an edge would disconnect the graph, a different node is chosen. This allows us to remove up to $N - 2$ edges, ultimately resulting in a tree graph with $N - 1$ edges. In each panel of Fig.~\ref{fig:auc_classification}, the leftmost yellow marker corresponds to the tree graph, and the rightmost yellow marker shows the result for the fully connected graph. As expected, removing edges from the Laman graph leads to a steep drop in performance, consistent with the fact that the Laman graph represents the minimal configuration required for local rigidity. A further reduction in connectivity leads to a loss of information. Interestingly, randomly adding edges to the Laman graph does not immediately improve performance. Instead, we observe a plateau, indicated by the gray shaded band in each panel of Fig.~\ref{fig:auc_classification}. The AUC starts to increase only after adding approximately an order of magnitude more edges. Eventually, the performance matches that of the fully connected graph. Instead, the blue markers show the AUC achieved using various {\it unique-$k$} graph constructions for $k = 3, 4, 5, 6, 7, 10, 15$. Across all classification tasks, we find that peak performance is reached with significantly fewer edges compared to the modified Laman graphs. This indicates that insights from graph theory may aid in both optimizing machine learning-based classifiers and interpreting their performance.

\bibliographystyle{JHEP}
\bibliography{main.bib}

\end{document}